\documentclass[11pt]{article}
\usepackage{fullpage}
\usepackage{amssymb}
\usepackage{amsthm}
\usepackage{amsmath}
\newtheorem{theorem}{Theorem}
\newtheorem{lemma}[theorem]{Lemma}
\newtheorem{corollary}[theorem]{Corollary}

\usepackage{xspace}
\usepackage{comment}
\usepackage{url}
\usepackage{cite}
\usepackage{times}
\usepackage{graphicx}

\newcommand{\alert}[1]{\typeout{ALERT: \the\inputlineno: #1}\textbf{[[[ #1 ]]]}}
\excludecomment{suppress}

\newcommand{\concept}[1]{\textbf{#1}}

\DeclareMathOperator{\interior}{int}

\DeclareMathOperator{\polylog}{polylog}

\newcommand{\N}{{\mathbb N}}
\newcommand{\Z}{{\mathbb Z}}
\newcommand{\R}{{\mathbb R}}
\newcommand{\Q}{{\mathbb Q}}
\newcommand{\zero}{0}

\newcommand{\etal}{{\it et al.}\xspace}

\newcommand{\jhat}{\hat{\jmath}}

\newcommand{\cF}{{\cal F}}

\newcommand{\elements}{E}             
\newcommand{\inputsymbols}{\Sigma}
\newcommand{\localoutputmap}{o}
\newcommand{\globaloutputmap}{O}
\newcommand{\protocol}{(\elements, \rightarrow, \inputsymbols, \localoutputmap)}

\newcommand{\configs}{\mathcal{C}}
\newcommand{\stable}{\mathcal{S}}
\newcommand{\unstable}{\mathcal{U}}

\newcommand{\predicate}{\psi}

\newcommand{\reaches}{\ensuremath{\stackrel{*}{\rightarrow}}}

\DeclareMathOperator{\Population}{Pop}

\newcommand{\classCOUNT}{\mathbf{COUNT}}
\newcommand{\classSLIN}{\mathbf{SLIN}}
\newcommand{\classMOD}{\mathbf{MOD}}
\newcommand{\classCoreMOD}{\mathbf{coreMOD}}

\title{The Computational Power of Population Protocols}
\author{Dana Angluin
\\Yale University 
\and James Aspnes\thanks{Supported in part by NSF grants
CNS-0305258 and CNS-0435201.}
\\Yale University
\and David Eisenstat
\\Princeton University 
\and Eric Ruppert\thanks{Supported in part by the Natural Sciences and
Engineering Research Council of Canada.}
\\York University
}

\begin{document}
\maketitle

\begin{abstract}

We consider the model of population protocols
introduced by Angluin~\etal~\cite{AADFP04},
in which anonymous finite-state
agents stably compute a predicate of the multiset of their
inputs via two-way interactions in the all-pairs
family of communication networks.
We prove that all predicates stably
computable in this model (and certain
generalizations of it) are semilinear, 
answering a central open question
about the power of the model.
Removing the assumption of two-way interaction,
we also consider several variants of the model
in which agents communicate by anonymous 
message-passing where 
the recipient of each message is
chosen by an adversary and the sender is not identified to the
recipient.
These one-way models are distinguished by whether messages are delivered
immediately or after a delay, whether a sender can record that it has
sent a message, and whether a recipient can queue incoming messages,
refusing to accept new messages until it has had a chance to send out
messages of its own.
We characterize the classes of predicates stably
computable in each of these one-way models using natural subclasses 
of the semilinear predicates.
 
\end{abstract}


\section{Introduction}
\label{sec:introduction}


In 2004, Angluin~\etal~\cite{AADFP04} proposed a new model of
distributed computation by very limited agents called a
population protocol.  In this model, finite-state agents
interact in pairs chosen by an adversary, with both agents updating
their state according to a joint transition function.  For each such
transition function, the resulting population protocol is said to
stably compute a predicate on the initial states of the agents
if, after sufficiently many interactions in a fair execution, 
all agents converge to
having the correct value of the predicate.  
Motivating scenarios include models of the propagation of trust
in populations of agents~\cite{DiamadiFischer01} and interactions
of passively mobile sensors~\cite{AADFP04,AADFP06}.
Similar models of pairwise interaction
have been used to study
the propagation of diseases~\cite{Bailey1975} and
rumors~\cite{DaleyKendall1965} in human populations
and
to justify the Chemical Master Equation~\cite{Gillespie92},
suggesting that the model of population protocols may be fundamental
in several fields.


Because the agents in a population protocol have
only a constant number of states, it is impossible for them to adopt
distinct identities, making them effectively anonymous.
An agent encountering another agent cannot tell in general
whether it has interacted with that agent before.
Despite these limitations, populations of such agents can
compute surprisingly powerful predicates on their initial states
under a reasonable global fairness condition.
When each agent may interact with every other agent,
any predicate over the counts of initial states definable in
Presburger arithmetic is computable~\cite{AADFP04,AADFP06}.
When each agent has only a bounded set of
neighbors with which it can interact, linear-space
computable predicates are computable~\cite{AACFJP05}.


In this paper we give exact characterizations of the
class of stably computable predicates in the all-pairs
interaction graph for the original population protocol
model and certain natural variants of it that we introduce
to model the restrictions of one-way communication.
Most of the results in this paper have appeared
in extended abstract form in the two conference
papers~\cite{AAE06,AAER05}.
In Section~\ref{section:summary} we also correct
an erroneous claim in ~\cite{AAER05}.

\subsection{Stably Computable Predicates Are Semilinear}


Angluin~\etal~\cite{AADFP04,AADFP06} showed that
various common predicates such as parity of the number of agents,
whether agents in some initial state $a$ outnumber agents
in another initial state $b$, and so forth, could be stably
computed by simple population protocols.  They further showed that
population protocols could in fact compute any semilinear
predicate, which are precisely those predicates definable in
first-order Presburger
arithmetic~\cite{Presburger1929}.
But it was not known whether there were other, stronger predicates
that could also be computed by a population protocol, at least in the
simplest case where every agent was allowed to interact with every other agent.
We show that this is not the case: that the semilinear predicates are
precisely the predicates that can be computed by a population protocol
if there is no restriction on which pairs of agents can interact with each
other.  This gives an exact characterization of the predicates stably
computable by population protocols, answering the major open question
of~\cite{AADFP04,AADFP06}.


These results
also answer an open question in~\cite{AACFJP05}: whether
requiring population protocols to deal with stabilizing inputs
(rather than assuming all inputs are
available at the start of computation)
reduces their computational power.
We show that the answer is no; in both cases, the
stably computable predicates are precisely the semilinear predicates.


The semilinearity theorem is proved for a generalization of the model
of population protocols, and applies to stable 
computation in other models that have the property
that a partitioned subpopulation can still run on its own (in the sense
that agents do not have any mechanism to detect if additional 
agents are present but not interacting).
Other models with this property include vector addition
systems~\cite{HP78}, some forms of Petri nets, and 1-cell catalytic
P-systems.
The relationship between semilinear sets and catalytic
P-systems, which can be represented as population protocols in which
all state-changing interactions occur between an unmodified catalyst
and at most one non-catalyst agent, has previously been considered by
Ibarra, Dang and Egecioglu~\cite{IDE04}, who show that the sets accepted by  1-cell
catalytic P-systems are semilinear.  Our results extend this fact to
predicates stably computable in this model, even without the catalytic
restriction.


Semilinearity is strongly tied to the notion of stable computation,
in which a correct and stable output configuration must always be reachable at
any step of the computation.
Mere reachability of a desired final configuration is not enough:
the reachability sets of population
protocols are not in general semilinear.
An example of this phenomenon may be derived 
from the construction
given by Hopcroft and Pansiot
of a non-semilinear reachability set for a six-dimensional
vector addition system \cite{HP78}.

\subsection{One-way communication}


We also introduce variants of the
population protocol model
that use forms of one-way communication
analogous to traditional asynchronous
message-passing models, and exactly characterize their
computational power in the all-pairs communication
graph in terms of
natural subclasses of the semilinear predicates.
In the one-way models, pairwise interactions are split into
separate send and receive events that each affect at most a single agent.
These models may better reflect communication
in the context of sensor networks, 
where radio communication
may not be bidirectional, even between nearby sensors.
Moreover, one-way
message-passing primitives may be easier to implement in practice.


For the one-way models, we consider the following attributes.
The sender may be allowed to change its state as a result
of sending a message (transmission models), or not (observation models).
The send and receive events for a message may occur simultaneously 
(immediate delivery models) 
or may be subject to a variable delay 
(delayed delivery and queued delivery models).
In the queued delivery model, a receiver may choose to postpone incoming
messages until it has had a chance to send a message of its
own; in the simpler delayed delivery models, 
the receiver does not have this option.
Thus, we consider five one-way models: immediate and delayed
observation, immediate and delayed transmission, and queued transmission.


The immediate models are perhaps best understood in terms of
interactions between agents, while the delayed and queued models
are better thought of in terms of messages sent from one agent
to another.
In the transmission models, the sender is aware that it has sent
a message, and may update its state accordingly.
In the observation models, the sender is unaware of being observed
by the receiver, and therefore does not update its state.
For example,
consider a passive radio frequency tag that does not update
its state in response to being read; this is an example
of immediate observation, in which the message (tag data)
is immediately delivered (read), but the state of the sender is unchanged.
A web page that updates a counter in response
to a visit can be thought of as an example of immediate transmission:
the message (web page contents) is immediately delivered (viewed),
and the web page is aware of the transaction and updates its state
accordingly.


We give exact characterizations of the classes of predicates that
can be stably computed in each of the one-way models we consider.
We show that the model of queued transmission 
stably computes exactly the semilinear predicates and is
equivalent in power to the standard population protocol 
model with two-way interactions.
The other models are strictly weaker, and can be characterized
in terms of natural subclasses of the semilinear predicates.
Weakest of all is the delayed observation model, which
allows only the detection of the presence or absence of each
possible input symbol.
Next is the immediate observation model, which allows
counting of input symbols up to an arbitrary constant limit.
The immediate and delayed transmission models
are equivalent to each other in power, and add the ability
to count input symbols modulo an arbitrary constant.
These results give us a precise and detailed understanding of the
capabilities of the one-way models.


\section{Related Work}
\label{section:related}

\subsection{Population Protocols and Related Models}
\label{section:populationprotocols-survey}

Stable computation by population protocols was introduced in~\cite{AADFP04,AADFP06}.
It was shown that all semilinear predicates are stably computable
by population protocols in the all-pairs interaction graph.
It was also shown that the all-pairs interaction graph has the
least computational power, in the sense that
it may be simulated in any other connected interaction graph.
This work was inspired by models of the propagation of trust
studied by Diamadi and Fischer~\cite{DiamadiFischer01}.
Related models of automata with a central finite control
and storage consisting of an unordered multiset of tokens
was studied in~\cite{AADFP03}.

Angluin~\etal~\cite{AADFP04,AADFP06} also introduced probabilistic population protocols,
in which a uniform random choice of pairs to interact replaces the fairness
condition; this enables quantification of the probability of error and
the number of interactions to convergence as a function of $n$, the
number of agents in the population.
It was shown that each semilinear predicate can be computed in an
expected $O(n^2 \log n)$ interactions, and that
a register machine with a constant number of registers
with $O(\log n)$ bits per register could be simulated with inverse
polynomial error probability and polynomial slowdown.
Recent work~\cite{AAE06fast} has given faster protocols for
these problems under the assumption that a unique leader is present
in the population: each semilinear predicate can be computed in
an expected $O(n \log^4 n)$ interactions, and an improved register
machine simulation can be done with inverse polynomial error
in $O(n \polylog(n))$ interactions per step.

The question of what properties of the interaction graph are
stably computable by population protocols was studied in~\cite{AACFJP05}.
It was shown that for any $d$, there is a population protocol
that organizes any connected interaction graph of maximum
degree at most $d$ into a linear memory of $\Theta(n)$ bits, which
is asymptotically optimal.
In the same paper,
the population protocol model was extended to stabilizing inputs, 
in which each
agent has an input that may change finitely many times over
the course of the computation before stabilizing to a final
value; this permits composition of protocols, in which the
stabilizing outputs of one protocol become the stabilizing inputs
of another protocol.
It was shown that all the semilinear predicates can be computed
with stabilizing inputs in the all-pairs interaction graph.
Thus, the present paper shows that the precisely the same predicates are
stably computable with or without stabilizing inputs in
the all-pairs interaction graph.

Self-stabilizing population protocols were studied
in~\cite{AAFJ05}, which gives self-stabilizing protocols for token 
circulation in a directed ring, directing an undirected ring, 
local addressing in
a degree bounded interaction graph, and leader election in a ring
with a constant bound $k$ on the smallest nondivisor of the ring size.
The question of resilience to crash faults and transient faults
was considered by Delporte-Gallet~\etal~\cite{DFGR06}, who give
a general method to transform a population protocol into one
that can withstand up to $c$ crash faults and $t$ transient faults.

\subsection{Comparison with Asynchronous Message-passing}
\label{asynchronous-message-passing}


In an asynchronous message-passing model, 
agents communicate by sending messages.  
An agent may spontaneously send a message at any time, which is
delivered to a recipient at some later time.  The recipient may
respond to the message by updating its state and possibly sending one
or more messages.  In the standard asynchronous model, senders can choose the
recipients of their messages, and recipients are aware of the
identities of the senders of messages they receive; however, in the population
protocol models we consider, these assumptions are dropped.


Agents in the population protocol models are assumed to be
finite-state.  Moreover, algorithms in this model are uniform:  the
description of the protocol cannot depend on the number of agents in
the system.
Together with a transition rule that depends only on
the states of the two interacting agents, these assumptions
naturally yield a model in which agents are effectively anonymous.
In this respect, the population protocol models
are weaker than a typical message-passing model,
where processes have identities.  
In addition, not only does a receiver
not learn the identity of the sender, but a sender cannot direct its
message to a particular receiver.  This is unusual even in anonymous
message-passing models, which typically assume that a process can use
some sort of local addressing to direct messages to specific
neighbors.  

 
The question of what computations can be performed in anonymous
systems, where processes start with the same state and the same
programming, has a long history in theoretical distributed computing.
Many early impossibility results such as~\cite{Ang80} assume both
anonymity and symmetry in the communication model, which limits what
can be done without some mechanism for
symmetry-breaking.  See~\cite{FR03} for a survey of many such
impossibility results.
More recent work targeted specifically at anonymity
has studied what problems are solvable in
message-passing systems under various assumptions about the initial
knowledge of the processes~\cite{BV99,BV01,Sak99}, or in anonymous
shared-memory
systems where the properties of the supplied shared objects can often
(but not always, depending on the details of the model) be
used to break symmetry and assign
identities~\cite{AspnesSS2002,AGM02,BuhrmanPSV05,ES94,GR05,JT90,KOP00,LP90,PPTV98,Ten90}.
This work has typically assumed few limits on the power of the
processes in the system
other than the symmetry imposed by the model.


Asynchronous message-passing systems may be vulnerable to a 
variety of failures,
including failures at processes such as crashes or Byzantine faults,
and failures in the message delivery system such as dropped or
duplicated messages.  In this paper we assume fault-free executions;
however, Delporte-Gallet~\etal~\cite{DFGR06} characterize
the power of population protocols in the presence of crash faults
and transient faults.  


Because message
delivery is asynchronous, making any sort of progress requires
adopting some kind of fairness condition to exclude executions in which
indefinitely-postponed delivery becomes equivalent to no delivery.
A minimal fairness condition might be that if some process 
sends a particular message $m$ infinitely often, 
then each other process receives the same message $m$ infinitely often.  
In Section~\ref{section-local-fairness},
we show that
even with unbounded states and message lengths,
this minimal fairness condition
provides only enough power to detect the presence or absence of
each possible input
because of the very strong anonymity
properties of the model.
Instead we adopt a
stronger global fairness condition derived from that used
in~\cite{AADFP04,AADFP06}, which is defined and further discussed in 
Section~\ref{section:unified}.


Considering the communication capabilities of the population
protocol models, the two-way population protocol model is stronger 
than a typical message-passing model: communication between 
two interacting agents is instantaneous and bidirectional.
Instantaneous communication is also a feature of the 
immediate transmission and immediate observation models.
The observation models are weaker than message-passing 
in that the sender is not aware of sending a message.
In the immediate and delayed transmission models, an agent
must be prepared to receive and react to a message at all times.
In the queued transmission model, which most closely approximates
asynchronous message-passing, an agent may postpone receiving
a message until it has sent messages of its own, and we show
that this capability is essential to achieve the full power of the model.


\section{Preliminaries}
\label{section:preliminaries}

Let $\N$ denote the set of natural numbers, $0, 1, 2, \ldots$.
The set of all functions from a set $X$ to a set $Y$ is denoted $Y^X$.
Let $E$ be a finite nonempty set. For all $f,g \in \mathbb{R}^E$,
we define the usual vector space operations
\begin{align*}
(f+g)(e) & := f(e)+g(e) \qquad \forall e \in E \\
(f-g)(e) & := f(e)-g(e) \qquad \forall e \in E \\
(c f)(e) & := c f(e) \qquad \forall c \in \mathbb{R}, e \in E \\
f \cdot g & := \sum_{e \in E} f(e)g(e).
\end{align*}
Abusing notation, we define a 0 vector and standard basis vectors
\begin{align*}
0(e) & := 0 \qquad \forall e \in E \\
e(e') & := [e=e'] \qquad \forall e,e' \in E,
\end{align*}
where $[condition]$ is $1$ if $condition$ is true and $0$ otherwise.
We define a natural partial order on $\mathbb{R}^E$ componentwise:
\begin{align*}
f \le g & \Leftrightarrow (\forall e \in E) f(e) \le g(e).
\end{align*}
Next we define the set of \concept{populations} on $E$:
\begin{align*}
\Population(E) & := \mathbb{N}^E \setminus \{0\}.
\end{align*}
These may be interpreted as the nonempty multisets on $E$:  for any $f\in
\Population(E)$ and $e\in E$, $f(e)$ represents the multiplicity of
the element $e$ in the multiset represented by $f$. 
Then, the partial order $\le$ corresponds to 
the subset order on multisets.


\section{A Unified Framework}
\label{section:unified}

In each of the models we consider, a \concept{protocol} can be considered
as determining five components:
a countable set $\configs$ of \concept{configurations};
a set of \concept{input symbols} $\inputsymbols$;
a binary relation $\rightarrow$ on $\configs$
that captures when the first configuration 
\concept{can reach} the second in one step;
a function $I:\Population(\inputsymbols)\rightarrow\configs$ that takes
inputs to \concept{initial} configurations;
and a partial function $\globaloutputmap:\configs\rightarrow\{0,1\}$
that gives the \concept{output} of each configuration on which it is defined.
We take $\reaches$ to be the reflexive-transitive closure of
$\rightarrow$.
We say $c'$ is \concept{reachable} from $c$ if $c\reaches c'$.
In this unified framework, we make the following definitions.


A configuration $c$ is \concept{output stable with output} $b$
if $\globaloutputmap(c) = b$ and for every $d$ such that $c \reaches
d$, $\globaloutputmap(d) = b$.
We define $\stable_b$ to be the set of configurations that are output stable
with output $b$, for $b \in \{0,1\}$, and define $\stable = \stable_0
\cup \stable_1$ to 
be the set of all \concept{output stable} configurations.
Thus a configuration is output stable if and only if its output is
defined, and every configuration reachable from it has the same
defined output.


An \concept{execution} is a (finite or infinite) sequence of configurations
$c_0,c_1,\ldots$ such that
for all $j$, we have $c_j \rightarrow c_{j+1}$.
An execution $c_0,c_1,\ldots$ is \concept{fair}
if for all $c' \in \configs$, either there exist infinitely many $j$
such that $c_j = c'$
or there exists $j$ such that $c_j \not\reaches c'$.
This is equivalent to the global fairness condition defined in~\cite{AADFP04,AADFP06},
and implies that a global configuration that is infinitely often
reachable in a fair execution must occur infinitely often in that execution.
This fairness condition may be viewed as an attempt to capture
useful probability $1$ properties in a probability-free model.
In Section~\ref{section-local-fairness} we show that a weaker, 
but plausible, fairness condition severely limits the power of the model.
Since $\configs$ is countable, 
any finite execution is a prefix of a fair execution, which can
be constructed as follows.
Fix an enumeration of $\configs$ where each configuration appears
infinitely often.  Then, starting with the finite execution, 
repeatedly extend it with a sequence of configurations that reaches the
next configuration in the enumeration that is reachable from the last
configuration of the execution constructed in the previous step.


A fair execution $c_0,c_1,\ldots$ \concept{converges with output $b$}
if there exists an $m$
such that for all $j \ge m$, the function $\globaloutputmap$
is defined on $c_j$ and $\globaloutputmap(c_j) = b$.
In particular, a fair execution converges with output $b$ if and only if
it reaches an output stable configuration with output $b$.
A protocol is \concept{well-specified} if,
for all initial configurations $c_0 \in I(\Population(\inputsymbols))$, 
there exists a $b$ such that all fair executions $c_0,c_1,\ldots$ converge
with output $b$.
In a well-specified protocol, every fair execution starting from an
input configuration converges with an output that is determined by
that input configuration.


A well-specified protocol induces a predicate
$\predicate : \Population(\inputsymbols) \rightarrow \{0,1\}$.
We say that this protocol \concept{stably computes} $\predicate$.
The results in this paper characterize the predicates $\predicate$
stably computable in various models of finite local state 
distributed computing.


\section{Definitions of Models}
\label{section:models}

In this section we define the standard model of two-way population protocols 
in the all-pairs interaction graph
and the one-way variants we consider, 
as well as the Abstract Model, which subsumes all of these.
For each model, we specify the configurations, 
input alphabet, one-step reachability relation,
input map and output map required by the unified framework above.


\subsection{Two-way}

A \concept{two-way} protocol is specified by five components:
$Q$, a finite set of states;
$\inputsymbols$, a finite set of input symbols;
$\delta : Q \times Q \rightarrow Q \times Q$, 
a joint transition function;
$\iota : \inputsymbols \rightarrow Q$, the initial state mapping;
and $\localoutputmap : Q \rightarrow \{0,1\}$, the individual output function.

We define
\begin{align*}
\configs & := \Population(Q)\\
I(x) & := \sum\limits_{\sigma\in\Sigma}x(\sigma)\iota(\sigma)\\
\globaloutputmap(c) & := b \textup{ if for all } q\in Q,\ c(q) \ge 1
\Rightarrow \localoutputmap(q) = b
\end{align*}
We define $c \rightarrow c'$ if
$q_1 + q_2 \le c$,
$c'=c-q_1-q_2+q_1'+q_2'$ and $\delta(q_1,q_2)=(q_1',q_2')$.


A configuration in this model is a multiset that gives the states of
all the agents.
Because agents do not have identifiers and we are considering
the all-pairs communication graph, agents in the same state
are interchangeable; thus, the multiset of their states
completely specifies the global state of the population.
An initial configuration sets the state of each agent
according to the input symbol assigned to it.
A step is an interaction between two agents 
and simultaneously
updates both of their states according to the value of the
joint transition function of their current states.
If the transition used is $\delta(q_1,q_2)$,
we refer to the agent in state $q_1$ as the \concept{initiator}
and the other agent as the \concept{responder}.

Each agent has an individual output of $0$ or $1$ determined
by its current state via the function $\localoutputmap$.
The configuration output function is $0$ (respectively, $1$)
if all the individual outputs are $0$ (respectively, $1$).
If the individual outputs are mixed $0$'s and $1$'s, then
the configuration output function is undefined.


\subsection{Transmission with Queuing}

A \concept{queued transmission} protocol is specified by seven components:
$Q$, a finite set of states;
$M$, a finite set of messages that is disjoint from $Q$;
$\inputsymbols$, a finite set of input symbols;
$\delta_s : Q \rightarrow M \times Q$,
a transition function for sent messages;
$\delta_r : Q \times M \rightarrow Q$,
a partial transition function for received messages;
$\iota : \inputsymbols \rightarrow Q$,
the initial state function;
and $\localoutputmap : Q \rightarrow \{0,1\}$,
the individual output function.
%
%
%
We define
\begin{align*}
\configs & := \{c \in \N^{Q\cup M} : c(q)>0 \textup{ for some } q\in Q\}\\
I(x) & := \sum\limits_{\sigma\in \Sigma}x(\sigma)\iota(\sigma)\\
\globaloutputmap(c) & := b \textup{ if for all }
	q\in Q,\:c(q) \ge 1 \Rightarrow \localoutputmap(q) = b.
\end{align*}
We have $c \rightarrow c'$ if 
for some state $q\in c$ and some message $m$, $c' = c-q+q'+m$ 
where $(m,q')=\delta_s(q)$;
or if there exists a message $m \in c$ and a state $q\in c$ such that
$c' = c-q-m+\delta_r(q,m)$.
In the latter case, $\delta_r(q,m)$ must be defined.


A configuration in this model is the multiset of all agents'
states and all messages in transit.  (They cannot be confused, because
$Q$ and $M$ are disjoint.)
An input configuration has no messages in transit
and sets the state of each agent
according to the input symbol assigned to it.
A step is either a send event, in which an agent in state
$q$ adds a message $m$ to the multiset of messages in transit
and goes to state $q'$ (according to $\delta_s(q) = (m,q')$,)
or a receive event, in which a message $m$ is removed from
the multiset of messages in transit and is delivered to an
agent in state $q$, which updates its state to $q'$ (according
to $\delta_r(q,m) = q'$.)
If $\delta_r(q,m)$ is not defined, then message $m$ cannot be
delivered to an agent in state $q$; this permits agents
to refuse to receive messages temporarily.
(We note that because an agent in an observation
model does not change state upon being observed,
there would be no way for it to leave a non-receive
enabled state; thus we do not consider a queued
observation model.)


\subsection{Immediate Transmission and Observation}

\concept{Immediate transmission}
is a special case of the two-way model in which
the state of the initiator
is updated independent of the state of the responder.
That is, there exist functions
$\delta_1 : Q \rightarrow Q$ and $\delta_2 : Q \times Q \rightarrow Q$
such that for all $q_1,q_2 \in Q$,
we have $\delta(q_1,q_2) = (\delta_1(q_1),\delta_2(q_1,q_2))$.
Thus, the initiator (or sender) is aware of the
fact that an interaction has taken place (or of sending
a message), and may update its state accordingly, but
it is not aware of the state of the responder (or receiver.)

\concept{Immediate observation}
is a special case of immediate
transmission in which $\delta_1$ is the identity function.
This is the situation in which the initiator (or sender)
is not aware of being ``observed'' by the responder
(or recipient) and therefore does not have an opportunity
to update its state.


\subsection{Delayed Transmission and Observation}

\concept{Delayed transmission}
is a special case of queued transmission,
with the requirement that $\delta_r$ be a total function.
In this case, the recipient does not have the option
of temporarily refusing to receive messages, which means that
it is in danger of being ``overwhelmed'' by incoming
messages.  Our characterization results show that
this indeed limits the power of protocols in this model.

\concept{Delayed observation}
is a special case of delayed
transmission in which for all $q \in Q$,
we have $\delta_s(q) = (q,m)$ for some $m \in M$.
In this model, the weakest of those we consider,
an agent can neither refuse incoming
messages nor update its state when it has sent
a message.


\subsection{The Abstract Model}

In order to present our semilinearity characterization
in full generality, we introduce
another model. 
All of the preceding single step rules
can be described in terms of replacing one collection
of elements (states or messages)
with another without regard to the other elements
in the configuration.
In this model it is also convenient to identify input symbols
with the states they map to, dispensing with the need for
an initial state map $\iota$.

$\elements$ is a set of elements, which may be either states or messages;
$\inputsymbols \subseteq \elements$ is a set of input symbols;
$\rightarrow$ is a relation on $\Population(\elements)$ such that if
$c\rightarrow c'$, then for all $d \in \Population(\elements)$,
we have $c+d \rightarrow c'+d$;
$\localoutputmap : \elements \rightarrow \{0,1\}$ is the individual
output map.
Then,
\begin{align*}
\configs & := \Population(\elements) \\
I(x) & := x, \textup{and}\\
\globaloutputmap(c) & := b \textup{ if for all } e\in E, c(e)\ge 1 \Rightarrow \localoutputmap(e)=b.
\end{align*}

To see that this generalizes the two-way model, we take $E$ to be the
disjoint union of the input symbols and states of the two-way model,
and treat the input symbol $\sigma$ as equivalent to the state
$\iota(\sigma)$, extending the individual output map $\localoutputmap$
to $\inputsymbols$ 
by $\localoutputmap(\sigma) = \localoutputmap(\iota(\sigma))$.
To see that the Abstract Model
also generalizes the queued transmission model,
we define $E$ to be the disjoint union of the input symbols,
states, and two copies of the messages of the queued transmission
model.
Having two copies of each message $m$ allows us to designate one
copy as having output $0$ and the other as having output $1$.
Again we treat input symbol $\sigma$ as equivalent to the state
$\iota(\sigma)$, and extend the individual output map $\localoutputmap$
to $\inputsymbols$ 
by $\localoutputmap(\sigma) = \localoutputmap(\iota(\sigma))$.
Also, we extend $\localoutputmap$ to the two copies of each message
by defining it to be the designated output of that copy.
To ensure that outputs propagate from agent states to messages,
we add rules that take a state of an agent and a copy of a message, and
change (if necessary) the designated
output of the message to be the same as
the output of the state.
Each send rule is modified to send a message with the same output
value as the sender, and the receive rules ignore the output values
of the messages.
These changes guarantee that when the outputs of the agents stabilize,
the designated outputs of the messages stabilize to the same thing.
Thus, every predicate stably computable in either the two-way or
the queued transmission model is stably computable in the Abstract
Model.


\subsection{Mirrors and Messages to Self}

Separating message transmission and receipt creates the
possibility that an agent may receive its own message.
Because senders are not identified, such an agent will
in general not be able to recognize the message as its own.  
This can be thought of as including
\concept{mirrors}, or self-loops in the interaction graph controlling which agents
can communicate, which we otherwise take to consist of all
ordered pairs of agents.
In general, we assume that this does not occur in the two-way and immediate
delivery models, which are perhaps best thought of as interaction models,
but may occur in the delayed and queued delivery models, on the
principle that once an anonymous message is sent
it may be delivered to anyone.
This has at most a minor
effect on the computational power of the models we consider,
which we note below as appropriate.


\section{Predicate Classes}

We now define the classes of predicates used in our characterizations.
Although stably computable predicates are defined only on 
$\Population(\inputsymbols)$,
our proofs are facilitated by defining predicate classes on $\Z^\inputsymbols$.
The \concept{support} of a predicate is the set of all inputs that
make it true.

\subsection{Semilinear Predicates}
\label{section:semilinearpredicates}

The most important class of predicates we consider is the class
of \concept{semilinear} predicates, which we write $\classSLIN$.
This class can be defined in several equivalent ways.

A \concept{linear set} is a set of the form
$\{b + k_1 p_1 + k_2 p_2 + \cdots + k_n p_n \mid k_1,k_2,\ldots,k_n \ge 0\}$,
where $b,p_1,p_2,\ldots,p_n$ are vectors.
The vector $b$ is the \concept{base} of the linear set, and the vectors
$p_i$ are the \concept{period} vectors.
A \concept{semilinear set} is a finite union of linear sets.
A \concept{(semi)linear predicate} is 
a predicate whose support is (semi)linear predicate.

\begin{figure}
\begin{center}
\includegraphics[scale=0.8]{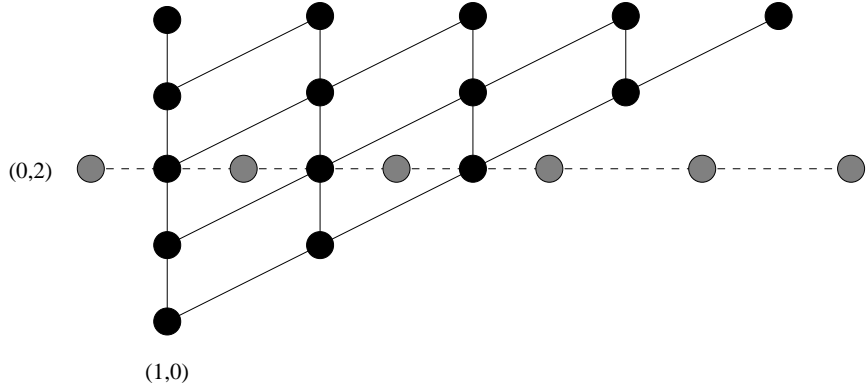}
\end{center}
\caption{A semilinear set $S$, equal to the union of the linear set
of all points
$\{ (1,0) + k_1 (1,0) + k_2 (0,2) \}$ (dark circles)
and the linear set
$\{ (0,2) + k_3 (2,0) \}$ (shaded circles).}
\label{fig-semilinear}
\end{figure}

Semilinear sets are also precisely the sets definable by first-order
formulas in \concept{Presburger arithmetic}~\cite{Presburger1929}
which are formulas in arithmetic that use only $<$, $+$, $0$, $1$, and the
standard logical quantifiers and connectives.  Here the
set consists of all satisfying assignments of the free variables; for
example, the semilinear set 
$S = \{ (1,0) + k_1 (1,0) + k_2 (0,2) \} \cup \{ (0,2) + k_3 (2,0)\}$,
depicted in Figure~\ref{fig-semilinear},
consists precisely of the satisfying assignments $(x,y)$ of the
formula
\begin{equation}
\label{eq-semilinear-example}
\left(\exists z: (z \ge 0) \wedge (x = z + z + 1) \wedge (y \ge z)\right)
\vee
\left(\exists z: (z \ge 0) \wedge (x = z + z) \wedge (y = 1 + 1)\right),
\end{equation}
where $x = y$ abbreviates $\neg ((x < y) \vee (y < x))$ and $x \ge y$
abbreviates $\neg (y < x)$.
It follows immediately from the correspondence between semilinear sets and
Presburger formulas that 
the semilinear sets are closed under complement, finite intersection
and finite union.
Thus a predicate is semilinear if and only if its complement
is semilinear.

A curious and useful property of Presburger formulas is that
all quantifiers (and their bound variables) can be eliminated by
the addition of binary relations 
$\equiv_m$ that test for equality modulo $m$ for any
nonnegative integer $m$~\cite{Presburger1929}.
For example, the formula (\ref{eq-semilinear-example})
defining $S$ can be rewritten
without quantifiers as
\begin{displaymath}
((x \ge 0) \wedge (x \equiv_2 1) \wedge (x \le y + y + 1))
\vee
((x \ge 0) \wedge (x \equiv_2 0) \wedge (y = 1 + 1)).
\end{displaymath}

This yields another characterization:
a semilinear predicate is a boolean combination of
threshold predicates,
whose support takes the form
$\{x \mid x \cdot v \ge r\}$ for some $v\in \Z^\inputsymbols$ and
$r\in \Z$,
and modulo predicates, whose support takes the form
$\{x \mid x \cdot v \equiv r \pmod m\}$ for some $v\in
\Z^\inputsymbols$ and $r,m\in \Z$ with $m > 0$.
Viewed geometrically, sets of the first type consist of points
on one side of a hyperplane, and sets of the second type are lattices.
As an example of a predicate of the first type, consider \concept{comparison},
which is true if the number of $a$'s in the input exceeds the number of
$b$'s in the input.
As an example of a predicate of the second type, consider \concept{parity},
which is true if the number of $a$'s in the input is odd.

In yet another characterization,
Parikh's Theorem~\cite{Parikh1966} 
shows that a subset $S$ of $\N^d$ is semilinear
if and only if there is a context-free language $L$ over an alphabet
of $d$ symbols such that $S$ consists of the vectors
of multiplicities of alphabet symbols of strings in $L$.
Moreover, the same statement holds with regular languages
in place of context-free languages.
Using this characterization, it is not difficult 
to see that the following predicates on the number
of $a$'s and $b$'s in the input are not semilinear:
the number of $a$'s is a prime,
the number of $a$'s is a square,
the number of $a$'s is a power of $2$,
and the number of $a$'s is bounded
above by $\sqrt{2}$ times the number of $b$'s.

\subsection{The Classes $\classMOD$ and $\classCoreMOD$}

We write $\classMOD$ for the class of boolean combinations 
of modulo predicates only.

Let $\inputsymbols' \subseteq \inputsymbols$ be any nonempty subset of
the input alphabet.
Then $x\in \Z^\inputsymbols$ 
is a \concept{$k$-rich profile with respect to} $\inputsymbols'$ if
$x(\sigma') \ge k$ for all $\sigma' \in \inputsymbols'$ and
$x(\sigma) = 0$ for all $\sigma \in \inputsymbols \setminus \inputsymbols'$.
Let $\predicate$ and $\predicate'$ be predicates on $\Z^\inputsymbols$.
If $\predicate(x) = \predicate'(x)$ for every $x$ that is a
$k$-rich profile with
respect to $\inputsymbols'$, then we say that
$\predicate$ and $\predicate'$ are \concept{$k$-similar with respect
to $\inputsymbols'$}.

The class $\classCoreMOD$ is the class of predicates $\predicate$ such that
for every nonempty $\inputsymbols' \subseteq \inputsymbols$,
there exists a $\predicate' \in \classMOD$ and $k \ge 0$
such that $\predicate$ is 
$k$-similar to $\predicate'$ with respect to $\inputsymbols'$.
Clearly $\classMOD \subseteq \classCoreMOD$, so the parity predicate
is in $\classCoreMOD$.
The comparison predicate is not in $\classCoreMOD$ because it is
not $k$-similar to any modulo predicate with respect to $\{a, b\}$ for
any $k \ge 0$.
However, if we consider the predicate over the alphabet $\{a, b, c \}$ that
is true when there is exactly one $c$ and the number of $a$'s exceeds the
number of $b$'s, then this predicate is in $\classCoreMOD$.
To see this, note that the predicate is false when the number of $c$'s
is $0$ or at least $2$, so $k = 2$ suffices as a witness for any
nonempty subalphabet.

\subsection{Simple Threshold Predicates}

We define a \concept{simple threshold predicate} to be a threshold
predicate with support $\{x \mid x \cdot v \ge r\}$
in which $v = \sigma$ for some input symbol $\sigma$.
An example of a simple threshold predicate is one that is true
when the number of $a$'s is at least $5$.
Then we define $\classCOUNT_k$ to be the class of
boolean combinations of simple threshold predicates in which
the threshold value $r \le k$.
A predicate in $\classCOUNT_k$ is completely
determined by the counts of the input symbols truncated at $k$.
For example, when $k = 1$, such a predicate depends only on the
presence or absence of each input symbol.
Finally, $\classCOUNT_*$ is the union of the classes $\classCOUNT_k$
for $k = 1,2,3, \ldots$.
The comparison predicate is an example of a threshold predicate that
is not in $\classCOUNT_*$.


\section{Summary of Characterizations}
\label{section:summary}

The computational power of the population protocol models we consider
is summarized in Figure~\ref{fig:power}.
For each model we give the class of predicates on $\Population(\inputsymbols)$
that can be stably computed by protocols in the model.
Protocols in the Abstract Model, which subsumes the two-way and queued transmission
models, stably compute exactly the semilinear predicates.
The immediate and delayed transmission models are equal in power,
and stably compute exactly those semilinear predicates that are 
in $\classCoreMOD$;
they cannot stably compute the comparison predicate.
Immediate observation protocols stably compute exactly those predicates
determined by the counts of the input symbols truncated at $k$ for some $k$;
they cannot stably compute the comparison predicate or the parity predicate.
Delayed observation protocols stably compute exactly those predicates
determined by the presence or absence of each input symbol; they
cannot stably compute the comparison predicate, the parity predicate
or simple threshold predicates for thresholds greater than $1$.


The results in~\cite{AAER05} incorrectly claimed
that any two-way stably computable predicates $k$-similar with
respect to $\inputsymbols$ to a predicate in $\classMOD$ is
stably computable in the immediate and delayed transmission models,
omitting the quantification over all nonempty subalphabets 
of the input alphabet.
To see why the more complex condition is required, consider
the predicate over the alphabet $\{a, b, c\}$ that is true if there is
{\it at most one} $c$ in the input and the number of $a$'s exceeds the
number of $b$'s.
This predicate is $2$-similar to the constant false predicate with
respect to $\{a, b, c\}$ but it is not in the class $\classCoreMOD$,
because for no $k \ge 0$ is it $k$-similar to a predicate in
$\classMOD$ with respect to the subalphabet $\{a, b\}$.
This predicate is not stably computable in the immediate and delayed
transmission models.

\begin{figure}[ht]
\centering
\leavevmode
\begin{tabular}{|c|c|}
\hline
Model & Power \\
\hline
Abstract & $\classSLIN$ \\
Two-way & $\classSLIN$ \\
Queued Transmission & $\classSLIN$ \\
Immediate Transmission & $\classSLIN \cap \classCoreMOD$ \\
Delayed Transmission & $\classSLIN \cap \classCoreMOD$ \\
Immediate Observation & $\classCOUNT_*$ \\
Delayed Observation & $\classCOUNT_1$ \\
\hline
\end{tabular}
\caption{The Power of Population Protocols}
\label{fig:power}
\end{figure}


\section{Protocols}
\label{section:protocols}

We first give protocols for each model in Figure~\ref{fig:power} to establish
that each model is at least as powerful as claimed.


\subsection{The Two-Way Model}
\label{section:two-way-protocols}

From~\cite{AADFP04,AADFP06} we have the following.
\begin{theorem}
\label{theorem:two-way-lower}
Every semilinear predicate on $\Population(\inputsymbols)$ 
is stably computable by a two-way population protocol.
\end{theorem}


To help us describe protocols in the one-way models,
we here give specific two-way protocols
for simple threshold, modulo and general threshold predicates.

\paragraph{Simple threshold predicates.}
The following protocol computes the simple threshold predicate
$\predicate(x) := [x \cdot \sigma \ge k]$, 
which is true when there are at least $k$ occurrences of the input symbol 
$\sigma$ in the input $x$.
\begin{align*}
Q & := \{0,1,\ldots,k\} \\
\delta(q_1,q_2) & := \begin{cases}
(q_1,k) & \textup{if } q_1 = k \\
(q_1,q_2+1) & \textup{if } 1\le q_1<k \textup{ and } q_1 = q_2 \\
(q_1,q_2) & \textup{otherwise}
\end{cases} \\
\iota(\sigma') & := [\sigma' = \sigma] \\
\localoutputmap(q) & := [q = k]
\end{align*}
Initially, agents with input $\sigma$ are in state $1$ and all other agents
are in state $0$.
As this protocol runs, the states of the agents make a ``tower''. 
All but one of the agents in state $i$ advance to state $i+1$ in each case. 
This tower will extend to $k$ if and only if there are initially
at least $k$ agents in nonzero states.
We note that this is an immediate observation protocol because the
state of the initiator remains unchanged by an interaction.


\paragraph{Active and passive agents.}
In the protocols for the modulo predicate $\predicate(x) = [x\cdot v
\equiv r \pmod m]$ and the threshold predicate $\predicate(x) =
[x\cdot v \ge r]$
agents fall into one of two categories:
active or passive. Every agent is initially active, 
and there is at least one active agent at all times.
Each active agent also has a data value, 
initially $\sigma \cdot v$, 
where $\sigma$ is the input symbol for this agent.
When a passive responder meets an active initiator, it
copies the output of the active initiator.
When two passive agents meet, nothing happens.
When two active agents meet, they attempt to combine their data values,
and one of them may become passive.


\paragraph{Modulo predicates.}
For a modulo predicate, $\predicate(x) = [x \cdot v \equiv r \pmod m]$,
the data values of active agents combine by sum modulo $m$.
The initiator becomes passive while the responder remains
active and keeps the combined data value.
Eventually, exactly one agent is active and has the correct sum modulo $m$,
and distributes the correct output to all of the other agents.
We note that this is an immediate transmission protocol because
the state of the initiator is updated uniformly, independent
of the state of the responder.


\paragraph{Threshold predicates.}
The situation for a threshold predicate,
$\predicate(x) = [x \cdot v \ge r]$, is more complicated.
In the end there may be multiple active agents, 
but they will all agree on the output.
We assume without loss of generality that $r \ge 0$.
The active states are a range of values 
that include all possible initial data values and $2r-1$.
The output is $1$ if and only if the data value is at least $r$.
Suppose $u$ and $v$ are the data values when two active
agents meet.
If $u+v$ is representable by an active state, one
agent remains active with this data value; the other becomes passive.
Otherwise, both agents remain active and they average their
data values, that is, one becomes
$\lceil \frac{u+v}{2} \rceil$ and the other becomes
$\lfloor \frac{u+v}{2} \rfloor$.
These transitions maintain the invariant that the sum of all
the active agents' data
values is $x\cdot v$.
Note that this protocol involves updates of both
states depending on both states;  
it is not an immediate transmission protocol.

To show correctness, we distinguish three cases based
on the sum of all the initial data values. 
In the first case, the sum is negative.
Eventually, no active agent will have a positive data value.
In the second case, the sum is between $0$ and $r-1$.
There will eventually be exactly one active agent with this data value.
In the third case, the sum is at least $r$. 
The number of active agents
with data value less than $r$ never increases.
The only way it can increase is after an
interaction involving an active agent with value at least $r$
where both agents remain active.
This will happen, however, only in the case when the two agents
average their values, which never happens with an agent with
value at least $r$ unless the result is both agents having value at least $r$.
After there are no more agents with negative values, 
any interaction with an agent whose value
is less than $r$ will decrease the number of such agents.


\subsection{Queued Transmission}

When combined with Theorem~\ref{theorem:two-way-lower},
the following theorem implies that every semilinear predicate on
$\Population(\inputsymbols)$
is stably computable in the queued transmission model.

\begin{theorem}
\label{theorem:queued-trans-sim}
Every predicate on $\Population(\inputsymbols)$
stably computable in the two-way model
is stably computable in the queued transmission model.
\end{theorem}

\begin{proof}
Given any protocol in the two-way model,
we describe a simulation of it by a protocol in the queued transmission model.
Each agent can store up to two
states from the two-way protocol. 
Initially, each agent has one state,
which is determined  from  the input symbol for that agent
by the initial state map of the protocol being simulated.
Agents transfer states by sending a message.
An agent called upon to send 
will send a message containing the state it has been holding longest. 
(If it is not currently holding any states, it sends a null
message.) 
Any agent with free space is eligible to receive a message.
Whenever an agent receives its second state, it uses the transition function
from the two-way protocol to have the two interact, 
with the state it already has acting as the initiator.

The simulated configuration cannot make any steps that were impossible under
the two-way model. 
Conversely, any sequence of interactions made under
the two-way model can be achieved by first having every agent launch
all states it holds, 
and then by repeating the following actions: deliver two states
to a particular agent, and then have the agent release both in messages.
Therefore, the simulation is faithful.
\end{proof}


\subsection{Immediate and Delayed Transmission}

\begin{theorem}
\label{theorem:immediatedelayedtrans-lower}
Every predicate on $\Population(\inputsymbols)$
in the class $\classSLIN \cap \classCoreMOD$ is stably
computable in the immediate transmission model and also in the
delayed transmission model.
\end{theorem}

\begin{proof}
We first note that both models can stably compute 
simple threshold predicates and predicates in $\classMOD$.
This is clear for immediate transmission because the
two-way protocols given in Section~\ref{section:two-way-protocols}
for simple threshold and modulo predicates are immediate
transmission protocols.
For the delayed transmission model, we specify that a sender sends its
state and becomes passive, preserving its previous output.
Passive messages are ignored by all receivers.
An active receiver combines
its data with the data from an active message, remaining active.
(The data values $u$ and $v$ are combined as $(u + v) \bmod m$ for a
mod predicate or $\min(k,u+v)$ for a simple threshold predicate.)
A passive receiver sets its
state equal to an active message, becoming active again.
In a fair execution eventually every configuration will
contain exactly one active agent or active message in transit,
and when all agents have received the message, 
they can compute the correct output value.

Suppose $\predicate \in \classSLIN \cap \classCoreMOD$.
For each nonempty subset $\inputsymbols'$ of 
the input alphabet $\inputsymbols$,
we describe a protocol that computes $\predicate$ assuming
that $\inputsymbols'$ is exactly the set of symbols present in the
input $x$.
We can ``and'' this protocol with one that verifies this assumption
(it is a boolean combination of simple threshold predicates, and
therefore stably computable in this model), and ``or'' together
the resulting protocols for each choice of $\inputsymbols'$,
and the result will stably compute $\predicate$.

Because $\predicate \in \classCoreMOD$, there is a predicate
$\predicate' \in \classMOD$ and an integer $k\ge 0$ such that $\predicate$
is $k$-similar to $\predicate'$ with respect to $\inputsymbols'$.
Then $\predicate'$ is stably computable by an immediate (or delayed)
transmission protocol, as is the predicate that is true if 
$x$ is $k$-rich with respect to $\inputsymbols'$,
so the conjunction of these predicates correctly computes $\predicate$
when there are at least $k$ occurrences of each input symbol
from $\inputsymbols'$ and no occurrences of any other symbol.

In the remaining cases
some $\sigma \in \inputsymbols'$ occurs between $1$ and $k-1$ times.
Because $\predicate \in \classSLIN$, it is stably computable by a two-way
protocol.
For each input symbol $\sigma \in \inputsymbols'$ 
we describe below how to adapt the simulation of the two-way protocol
in the proof of Theorem~\ref{theorem:queued-trans-sim}
to show that the predicate that
is the conjunction of $\predicate$ and the predicate that is
true when there are between $1$ and $k-1$ 
occurrences of $\sigma$ in the input is stably computable in
the immediate (or delayed) transmission models.
Then we ``or'' these protocols together for all choices of
$\sigma \in \inputsymbols'$. 

The simulation of a two-way protocol by a queued transmission protocol
requires queuing in order to
prevent agents' storage from being overwhelmed by messages.
However, 
suppose it may be assumed that at least one and
at most $k-1$ copies of $\sigma$ can appear
in the input. 
In this simulation, each agent can store up to $k$ states of the
original protocol.
We give the (at most $k-1$)
agents receiving input $\sigma$ ``tokens'' and require
that every agent ``pay'' a token to the recipient every time
it transfers a state.
The result is that no agent can collect more than $k$ states: one for
the state it started with and $k-1$ transfers. At that point, the agent
has all of the tokens and cannot gain any more states.
Thus, an agent never runs out of space for the messages it receives.

To see that any step of the original protocol can be simulated in this
protocol,
consider a simulated configuration with two agents in states $q_1$ and
$q_2$.  First, the agent holding $q_1$ sends $q_1$ in a message.  (If
the agent does not have enough tokens to do so, deliver a message to
it so that it gets the necessary token.)  When $q_1$ is a message in
transit, some agent holds no states.  Deliver $q_1$ to such an empty agent.
Then, the agent holding $q_2$ sends $q_2$ in a message.  (Again,
deliver a token to this agent if required to make this happen.)
Deliver $q_2$ to the agent that holds $q_1$ to cause an
interaction between $q_1$ and $q_2$.
\end{proof}

Note in particular that immediate and delayed transmission protocols
can stably compute the predicate over the alphabet $\{a, b, c\}$ that
is true if there is exactly one $c$ and the number of $a$'s exceeds
the number of $b$'s.
Intuitively, the single $c$ in accepted inputs can be used to implement
a kind of flow control on agents' incoming messages.
The characterization in Section~\ref{section:immdeltrans-upper} shows
that the standard comparison predicate over $\{a, b\}$, which is true
if the number of $a$'s exceeds the number of $b$'s, is not stably
computable by immediate or delayed transmission protocols,
essentially because no such control of incoming messages is possible
in this case.


\subsection{Immediate Observation}

\begin{theorem}
\label{theorem:immediateobs-lower}
Every predicate on $\Population(\inputsymbols)$ in $\classCOUNT_*$ 
is stably computable in the immediate observation model.
\end{theorem}

\begin{proof}
Every predicate in $\classCOUNT_*$ can be expressed as a boolean combination of
simple threshold predicates that depend only on the count of one input symbol.
The two-way protocol in Section~\ref{section:two-way-protocols} for the simple
threshold predicate $[x \cdot \sigma \ge k]$ is in fact an immediate observation
protocol because only the responder's state is updated in each case.
\end{proof}

The protocol from Section~\ref{section:two-way-protocols}
uses $k+1$ states, which can
be reduced to $k$ (by removing state $0$) if the
input alphabet is unary.
It can be shown that $k-1$ states are not sufficient
to stably compute this predicate in the immediate observation model
in the case of a unary alphabet;
the proof is rather involved and will appear elsewhere.


\subsection{Delayed Observation}

\begin{theorem}
\label{theorem:delayedobs-lower}
Every predicate on $\Population(\inputsymbols)$
in $\classCOUNT_1$ is stably computable 
in the delayed observation model.
\end{theorem}

\begin{proof}
The two-way simple threshold protocol in
Section~\ref{section:two-way-protocols}
is also a valid delayed observation protocol
for $k=1$. However, because observation is delayed, an agent
may repeatedly advance its state after receiving a message from
itself, which means that the
protocol fails to compute the correct predicate in the delayed
observation model when $k \ge 2$.
\end{proof}

We observe that if agents do not receive messages from themselves,
then predicates in $\classCOUNT_2$ are stably computable in
the delayed observation model.


\section{Characterization of the Power of the Abstract Model}
\label{section:char-abstract}

To complete the characterizations in Figure~\ref{fig:power}, we have
to demonstrate that each model is limited to the indicated power.
In Section~\ref{section:one-way-char}, we consider the one-way models.
Here we consider the Abstract Model, which includes the
two-way model and the queued transmission model as special cases,
and prove the following semilinearity theorem.


\begin{theorem}
\label{theorem:main-semilinear}
Every predicate on $\Population(\inputsymbols)$ 
that is stably computable in the
Abstract Model is semilinear.
\end{theorem}


Because the two-way and queued transmission models
are special cases of the Abstract Model, and both models can stably
compute all the semilinear predicates over $\Population(\inputsymbols)$,
we have the following characterizations.

\begin{corollary}
\label{corollary:two-way-queued-char}
The predicates stably computable in
(1) the two-way model of population protocols,
or (2) the queued transmission model of population protocols
are exactly the semilinear predicates on $\Population(\inputsymbols)$.
\end{corollary}


Stable computation of a predicate by a two-way
population protocol with stabilizing inputs was defined in~\cite{AACFJP05}.
We do not repeat the definition here, but the idea is that each
agent has an input register that may change finitely many times over
the course of the execution before stabilizing to its final value,
and the goal is to compute a predicate on the multiset of all the
agents' final input values.
Because this is a more restrictive definition, every predicate 
stably computable by a two-way protocol with stabilizing inputs
is stably computable by a two-way protocol with fixed inputs,
but whether the converse held was left as an open problem.
However, in~\cite{AACFJP05} it was shown that every semilinear predicate 
on $\Population(\inputsymbols)$
is stably computable by a two-way protocol with stabilizing
inputs.
Thus we get the following corollary
showing that stabilizing inputs do not reduce the power of the 
standard two-way model.

\begin{corollary}
\label{corollary:stabilizinginputs-char}
The predicates stably computable by two-way
population protocols with
stabilizing inputs are exactly the semilinear predicates on
$\Population(\inputsymbols)$.
\end{corollary}


To prove Theorem~\ref{theorem:main-semilinear}, 
we must show that the support of any
stably-computable predicate is a semilinear set: in particular,
that there is a finite set of base points each attached to a
finitely-generated cone such that the support is
precisely the elements of these cones.  The first step, which 
requires the development of the machinery of
Section~\ref{section-groundwork} and is completed in
Section~\ref{section-pumping-lemma}, is to show that the support
can be decomposed into a finite collection of monoid cosets
that are \emph{not necessarily finitely generated}.  We then proceed,
in
Sections~\ref{section-big-proof-outline}
and~\ref{section-big-proof-details},
to show that any such
decomposition can be further decomposed into a finite covering by 
cosets of finitely generated monoids, which gives us the full result.


\section{Groundwork}
\label{section-groundwork}

We assume that $\predicate$ is a predicate stably computed by a protocol
$\protocol$ in the Abstract Model
and establish some basic results.  These will lead up to
the Pumping Lemma of Section~\ref{section-pumping-lemma}.


\subsection{Monoids, Groups, and Semilinearity}
\label{section-preliminaries-algebra}


A subset $M$ of $\Z^d$ is a \concept{monoid} if it contains the zero
and is closed under addition; if it is also closed under subtraction,
$M$ is a \concept{group}.
A monoid $M \subseteq \Z^d$ is \concept{finitely generated} if there
exists a finite subset $A \subseteq M$ such that every element of $M$
is a sum of elements from $A$.
It is a classic result in abstract algebra that every subgroup of $\Z^d$ 
is finitely generated~\cite{lang:algebra}, but 
submonoids are not always finitely generated. 
For example, the following monoid is not finitely generated.
\[M_{\sqrt{2}} = \{(i,j) \in \N^2: i \le \sqrt{2}j \}.\]
A subset $H$ of $\Z^d$ is a \concept{group coset} (resp., 
\concept{monoid coset}) 
if there exists an element $v \in \Z^d$ such that $H = v + G$ and $G$ 
is a group (resp., monoid).

Then from the definitions in Section~\ref{section:semilinearpredicates},
we have yet another characterization of linear and semilinear sets.
A subset $L$ of $\Z^d$ is \concept{linear} if it is a coset of a finitely
generated monoid in $\Z^d$, and is \concept{semilinear} if it is a finite
union of linear sets.


\subsection{Higman's Lemma}
\label{section-Higman}

We will make extensive use of some corollaries to Higman's
Lemma~\cite{Higman1952}, a fundamental tool in well-quasi-order theory.


\begin{lemma}
\label{lemma-minimal-elements}
Every subset of $\N^d$ under the inclusion ordering $\le$ has
finitely many minimal elements.
\end{lemma}


\begin{lemma}
\label{lemma-infinite-chain}
Every infinite subset of $\N^d$ contains an infinite chain (i.e., an
infinite totally ordered sequence).
\end{lemma}

These both follow from the fact that Higman's Lemma implies that $\N^d$ is a
\concept{well-quasi-order}, that is, a set in which any infinite sequence
$a_1, a_2, \ldots$ contains elements $a_i, a_j$ with $i < j$ and $a_i
\le a_j$.  The special case of Higman's Lemma given in
Lemma~\ref{lemma-minimal-elements} was proved earlier by Dickson
\cite{Dickson13}.


\subsection{Truncation Maps and Their Properties}
\label{section-truncation}


For each $k \ge 1$, we define a map $\tau_k$ from $\configs$ to $\configs$ by
\[\tau_k(c)(e) := \min(k, c(e)) \textup{ for all } e \in \elements. \]
This map truncates each component 
of its input to be at most $k$; 
clearly $\tau_k(c) \le c$ for all $c \in \configs$.
Two useful properties of $\tau_k$ are that it respects both inclusion 
and addition.


\begin{lemma}
\label{lemma:trunc-order}
For all $c, d \in \configs$ and $k \ge 1$, 
if $c \le d$ then $\tau_k(c) \le \tau_k(d)$.
\end{lemma}

\begin{proof}
For each $e \in \elements$, we have $c(e) \le d(e)$, 
so $\min(k,c(e)) \le \min(k,d(e))$.
Thus $\tau_k(c) \le \tau_k(d)$.
\end{proof}


\begin{lemma}
\label{lemma:trunc-add}
For all $c, c', d \in \configs$ and $k \ge 1$, 
if $\tau_k(c) = \tau_k(c')$, then $\tau_k(c+d) = \tau_k(c'+d)$.
\end{lemma}

\begin{proof}
For each $e \in \elements$, either $c(e) = c'(e)$ or both are at least $k$.
In either case, $\min(k, c(e)+d(e)) = \min(k, c'(e)+d(e))$, so
$\tau_k(c+d) = \tau_k(c'+d)$.
\end{proof}

\subsection{Truncation and Stability}
\label{section-truncation-stability}


Truncation is important because membership of a 
configuration $c$ in the set of output stable configurations
$\stable$ can be determined from a truncate of fixed size. 
Let $\unstable := \configs \setminus \stable$, 
the set of \concept{output unstable} configurations.


\begin{lemma}
\label{lemma:Uclosed}
For all $c \le d$, if $c \in \unstable$, 
then $d \in \unstable$ ($\unstable$ is closed upward under inclusion).
\end{lemma}

\begin{proof}
Suppose $c \in \unstable$ and $c \le d$.
Then either (1) $\globaloutputmap(c)$ is undefined, or 
(2) $\globaloutputmap(c) = b$, but for some configuration
$c'$ such that $c \reaches c'$ either $\globaloutputmap(c')$ is undefined
or $\globaloutputmap(c') \not= b$.
In case (1), $\globaloutputmap(d)$ is undefined and $d \in \unstable$. 
In case (2), $d = d-c+c \reaches d-c+c'$.
If $\globaloutputmap(d)$ is not defined, then $d \in \unstable$.
If $\globaloutputmap(c')$ is not defined, then $\globaloutputmap(d-c+c')$ is
not defined, and $d \in \unstable$.
If both $\globaloutputmap(d)$ and $\globaloutputmap(c')$ are defined, we
have that $\globaloutputmap(d) = \globaloutputmap(c) \not=
\globaloutputmap(c')$, and $\globaloutputmap(d-c+c')$ is either
undefined or equal to $\globaloutputmap(c')$,
so $d \in \unstable$.
Thus $\unstable$ is closed upwards under inclusion.
\end{proof}


\begin{lemma}
\label{lemma:trunc-upwards}
There exists $k \ge 1$ such that $c \in \unstable$ if and only if
$\tau_k(c) \in \unstable$.
\end{lemma}

\begin{proof}
By Higman's Lemma,
only finitely many elements $u_1,\ldots,u_n$ 
are minimal in $\unstable$, and because $\unstable$ is upwards closed, 
$c \in \unstable$ if and only if $u_i \le c$ for some $i$.
Let $k$ be the maximum value $u_i(e)$ for all $i \in \{1,\ldots,n\}$ 
and all $e \in \elements$.
Then $\tau_k(u_i) = u_i$ for each $i$.

Suppose $c \in \unstable$.
Then $u_i \le c$ for some $i$, so $u_i = \tau_k(u_i) \le \tau_k(c)$ 
by Lemma~\ref{lemma:trunc-order}, and thus $\tau_k(c) \in \unstable$.
Conversely, if $\tau_k(c) \in \unstable$, then $u_i \le \tau_k(c) \le c$ for some 
$i$, and therefore $c \in \unstable$.
\end{proof}


\begin{lemma}
\label{lemma:trunc-stable}
There exists $k \ge 1$ such that 
for all $c \in \configs$ and $b \in \{0, 1\}$, we have 
$c \in \stable_b$ if and only if $\tau_k(c) \in \stable_b$.
\end{lemma}

\begin{proof}
By Lemma~\ref{lemma:trunc-upwards}, there exists
$k \ge 1$ such that for all $c \in \configs$, we have
$c \in \unstable$ if and only if $\tau_k(c) \in \unstable$.
Taking the contrapositive, we have
$c \in \stable$ if and only if $\tau_k(c) \in \stable$.
Since truncation does not affect output, the conclusion follows.
\end{proof}


\subsection{Extensions}
\label{section-extensions}


We define a map $X$ from $\configs$ to 
subsets of $\N^{\inputsymbols}$ as follows.
\[X(c) := \{x \in \N^{\inputsymbols} \mid \textup{there exists }
d \ge c \textup{ such that }
c+x \reaches d \textup{ and } \tau_{k}(c) = \tau_{k}(d) \},\]
where $k$ is the constant from the conclusion of 
Lemma~\ref{lemma:trunc-stable}. If $c \in \stable$, then $X(c)$ 
is the set of inputs by which $c$ can be pumped.
We call such inputs the \concept{extensions} of $c$.
We first prove that pumping does not affect stable output.


\begin{lemma}
\label{lemma:Xpredicate}
If $x \in \Population(\inputsymbols)$ and 
$c \in \stable$ and $x \reaches c$, 
then $\predicate$ is constant on $x + X(c)$.
\end{lemma}

\begin{proof}
If $y \in X(c)$, then there exists $d \in \configs$ such that
$c + y \reaches d$ and $\tau_{k}(c) = \tau_{k}(d)$.
Since $c \in \stable$, by Lemma~\ref{lemma:trunc-stable} we have
$d \in \stable$, and $\globaloutputmap(c) = \globaloutputmap(d)$.
Thus $\predicate(x) = \predicate(x+y)$.
\end{proof}


We now prove that pumping operations can be composed, i.e., that 
$X(c)$ is a monoid.
\begin{lemma}
\label{lemma:Xmonoid}
$X(c)$ is a monoid for all $c \in \configs$.
\end{lemma}

\begin{proof}
We have $\zero \in X(c)$, with $d = c$ as a witness.
If $x_1, x_2 \in X(c)$, then there exist $d_1, d_2$ such that
$c \le d_1$ and $c \le d_2$ and $\tau_{k}(c) = \tau_{k}(d_1) = \tau_{k}(d_2)$ 
and $c + x_1 \reaches d_1$ and $c + x_2 \reaches d_2$.
Thus 
\[c + x_1 + x_2 \reaches d_1 + x_2 = (d_1 - c) + c + x_2 
\reaches (d_1 - c) + d_2.\]
Taking $d := d_1 + d_2 - c$, we have
$c \le d$ and $c + x_1 + x_2 \reaches d$ and 
$\tau_{k}(c) = \tau_{k}(d_2) = \tau_{k}(c + d_2 - c)$ 
and by Lemma~\ref{lemma:trunc-add}, 
$\tau_{k}(c + d_2 - c) = \tau_{k}(d_1 + d_2 - c) = \tau_{k}(d)$, 
since $\tau_{k}(c) = \tau_{k}(d_1)$. 
We conclude that $x_1 + x_2 \in X(c)$.
\end{proof}


\section{A Pumping Lemma for Stably Computable Predicates}
\label{section-pumping-lemma}
Given a set of inputs $Y \subseteq \Population(\inputsymbols)$, 
a \concept{monoid-coset covering of $Y$ with respect
to $\predicate$} is a set $\{(x_i, M_i)\}_{i\in I}$ 
of pairs of inputs and submonoids of $\N^\inputsymbols$ 
such that $Y \subseteq \bigcup_{i\in I} (x_i + M_i)$ 
and for all $i \in I$, we have
$x_i \in Y$ and 
$\predicate(x_i + M_i) = \{\predicate(x_i)\}$. 
We say $\predicate$ \concept{admits finite coset coverings} if 
for all $Y$ there exists a finite monoid-coset covering of $Y$ with respect 
to $\predicate$. 
The following lemma states that every stably computable predicate admits a 
finite monoid-coset covering. 
We show later (Theorem~\ref{theorem:general}) 
that any predicate that admits finite coset coverings 
is semilinear. 

\begin{lemma}
\label{lemma:pumping}
The (stably computable) predicate $\predicate$ admits finite coset coverings.
\end{lemma}

\begin{proof}
Consider any $Y\subseteq \Population(\inputsymbols)$.
If $Y$ is a finite set $\{x_1, \ldots, x_n\}$ then it has a trivial
finite covering $\{(x_i,\emptyset)\}_{1\leq i\leq n}$.
So assume $Y$
is infinite. 
Let $y_1, y_2, \ldots$ be any enumeration of $Y$
such that $y_i \le y_j$ implies $i \le j$.
(For example, 
fix an ordering of $\inputsymbols$.  Then enumerate the elements of
$Y$ ordered by $\le$, breaking ties according to lexicographic order.)
We define a family of sets 
$B_i \subseteq \Population(\inputsymbols) \times \stable$ 
inductively as follows.
\begin{align*}
B_0 & := \emptyset. \\
B_i & := \begin{cases} B_{i-1} \textup{ if there exists } (x,c) \in B_{i-1} 
\textup{ such that } y_i \in x + X(c) \\
B_{i-1} \cup \{(y_i,s(y_i))\} \cup \{(y_i,s(c+y_i-x)) \mid 
(x,c) \in B_{i-1} \textup{ and } x \le y_i\} \textup{ otherwise,}
\end{cases}
\end{align*}
where $s(d) \in \stable$ is any stable configuration reachable from $d$.
It is easy to see by induction on $i$ that, for all $(x,c) \in B_i$,
$x\reaches c$, and
in the construction, $s(d)$ is applied only to reachable
configurations $d$, so the existence of $s(d)$ is guaranteed
by the requirements of stably computing a predicate.  
So, Lemma \ref{lemma:Xpredicate} implies that $\psi$ is
constant on $x+X(c)$.
Let $B:= \bigcup_{i\ge 1} B_i$.
It follows from \ref{lemma:Xmonoid} that 
$\{(x, X(c)) \mid (x,c) \in B\}$ is a monoid-coset covering of $Y$ with
respect to $\predicate$.
We now show that $B$ is finite to prove that this is a finite covering.

Assuming to the contrary that $B$ is infinite, infinitely many different 
elements of $Y$ appear as first components of elements of $B$, since
each $B_i$ is clearly finite.
By Higman's Lemma, there exists an infinite chain $z_1 < z_2 < \ldots$ 
of such elements. Our construction guarantees the existence of 
associated configurations 
$\{d_i\}_{i\ge 1}$ such that $(z_i, d_i) \in B$ and 
$d_i + (z_{i+1} - z_i) \reaches d_{i+1}$ for all $i \ge 1$.

By Higman's Lemma again,
there exists an increasing function $f$ such that 
the sequences $(z_{f(i)})_{i\ge 1}$ and $(d_{f(i)})_{i\ge 1}$ 
are nondecreasing. Thus the sequence $\tau_{k}(d_{f(i)})$ 
reaches a maximum and becomes constant at some index $i = j$.
Consequently, we have $z_{f(j+1)} - z_{f(j)} \in X(d_{f(j)})$,
which contradicts the membership of $(z_{f(j+1)}, d_{f(j+1)})$ in $B$.
\end{proof}


Applying this lemma with $Y = \predicate^{-1}(1)$
(the support of $\predicate$)
we obtain a finite family of monoid cosets $x_i + M_i$ such that
\[\predicate^{-1}(1) = \bigcup_{i} (x_i + M_i).\]
This does not prove semilinearity by itself, since some monoid
$M_i$ might not be finitely generated.
However, every submonoid of $\Z^1$ is finitely generated, so
in the special case of a \emph{unary} alphabet, we have
already that $\predicate$ is semilinear, and therefore regular.


\begin{corollary}
\label{corollary:unary}
Every stably computable predicate over a unary alphabet is semilinear.
Thus, over a unary alphabet,
the stably computable predicates are
exactly the semilinear (in fact, regular) predicates.
\end{corollary}


For example, the unary predicate that is true if the number
of input symbols is a power of $2$ (or a prime, or any other non-regular
predicate) is not stably computable.
Another easy corollary suffices to show certain other predicates over
non-unary alphabets are not stably computable.


\begin{corollary}
\label{corollary:pumping}
Suppose $\predicate$ is a stably computable predicate
such that $L = \predicate^{-1}(1)$ is infinite.
Then $L$ contains an infinite linear subset.
\end{corollary}

\begin{proof}
By Lemma~\ref{lemma:pumping} there is a finite monoid-coset covering of $L$.
If $L$ is infinite, some $(x + M) \subseteq L$ in the covering must be infinite.
\end{proof}


As an application, consider
the set of inputs over the alphabet $\{a, b\}$
such that the number $b$'s
is the square of the number of $a$'s.
This is an infinite set with no infinite linear subset,
and is therefore not stably computable.
Using closure results for stably computable predicates
we can then show that
the set of all inputs over alphabet $\{a, b, c\}$ such
that the number of $c$'s is the product of the number of $a$'s
and the number of $b$'s is not stably computable.
The existence of a pumping lemma and the negative results
for these particular predicates were conjectured in \cite{AADFP04,AADFP06}.


\section{Proof of the Semilinearity Theorem: Outline}
\label{section-big-proof-outline}

We are now in a position to give an overview of how we go from the pumping
lemma (Lemma~\ref{lemma:pumping})
of Section~\ref{section-pumping-lemma} to the 
semilinearity theorem.

Recall the following set of points in $\N^2$.
\[M_{\sqrt{2}} = \{(i,j): i \le \sqrt{2}j\}.\]
This is a monoid but is not stably computable.
To see this, suppose the contrary.
By the Pumping Lemma, there exist 
monoid cosets $x_i+M_i$
for $1 \le i \le m$ such that
\[M_{\sqrt{2}} = \bigcup_{1 \le i \le m} (x_i + M_i).\]
Let $v = (-1, \sqrt{2})$; then $x \cdot v > 0$
for all $x \in M_{\sqrt{2}}$. 
Let $\epsilon = \min_i \{x_i \cdot v\}$.
Choose some $y \in M_{\sqrt{2}}$ such that $0 < y \cdot v < \epsilon$.
Then for some $i$, $y \in x_i + M_i$,
and $(y - x_i) \cdot v =y\cdot v-x_i\cdot v <\epsilon-\epsilon = 0$.
Thus for a sufficiently large $n \in \N$,
$(x_i + n(y - x_i)) \cdot v < 0$, which contradicts
the fact that $x_i + M_i$ is a subset of
$M_{\sqrt{2}}$.
The issue here is that the line dividing the positive
and negative inputs cannot have an irrational slope
if the predicate is stably computable.
One ingredient of our proof is a generalization of this
idea to separating hyperplanes.

However, to be able to use separating hyperplanes,
we first must deal with separating ``intermixed'' positive and
negative inputs using their images in a finite group.
For example, consider the following set of points in $\N^2$, 
which is stably computable.
\[L = \{(i,j): i < j, (i+j) \mbox{ is odd}\}.\]
By first separating the points in $\N^2$ by their images
$(i \bmod 2, j \bmod 2)$ in the group $\Z_2 \times \Z_2$,
we get four subproblems in which lines suffice
to separate the positive and negative points.

A further issue is that because of
the relationship between the monoids $X(c)$ and
their cosets $x + X(c)$, instead of a single separating
hyperplane we have to consider finite sets of parallel
hyperplanes, which themselves may contain points from
the support of $\predicate$.
These points can be handled by induction on dimension,
but this 
requires that the main theorem be generalized to consider
the intersection of the support of $\predicate$ and
an arbitrary group coset (where we view each hyperplane
as a group coset.)
We shall prove the following theorem
by induction on the dimension of the group $G$; it clearly holds
when $G$ has dimension $0$, i.e., when $G$ is the trivial group.

\begin{theorem}
\label{theorem:general}
If $\predicate$ admits finite coset coverings then
for any group coset $H = x_0 + G \subseteq \Z^{\inputsymbols}$,
we have $\predicate^{-1}(1) \cap H$ is semilinear.
\end{theorem}

Note that together with Lemma~\ref{lemma:pumping}
this implies Theorem~\ref{theorem:main-semilinear}, as we can take
$H = \Z^{\inputsymbols}$.

The details of the proof are quite involved and are in
Section~\ref{section-big-proof-details}.
To summarize briefly, the overall strategy is:

\begin{enumerate}
\item By dividing the space into residue classes with respect to
appropriately chosen moduli, we can arrange for the vectors from 
the monoids associated with the cover 
to appear in all-positive and all-negative regions separated
by hyperplanes.  In this part of the proof we extend
$\N^\inputsymbols$ to $\Z^\inputsymbols$ to make use of the fact
that all subgroups of $\Z^\inputsymbols$ are finitely generated.
(Section~\ref{section:intermixed}.)
\item We then further map the problem from $\Z^\inputsymbols$ to
$\R^\inputsymbols$, and use techniques from convex geometry to show
that appropriate hyperplanes separating the monoids indeed exist.
(Section~\ref{section:hyperplanes}.)
We obtain a (looser) separation by \concept{slabs} (the space between 
two parallel hyperplanes)
on the inputs by observing that 
each input is displaced a uniformly bounded amount from its corresponding 
extension vector.
\item Moving to $\R^\inputsymbols$ allows for the possibility of
separating hyperplanes with irrational coefficients.  
Applying the Pumping Lemma as described above, we show that the
resulting separating hyperplanes are normal to a vector with rational 
coordinates and thus
correspond to cosets of subgroups of $\Z^\inputsymbols$.
(Section~\ref{section:rationality}.)
\item At this stage, we have shown that the positive inputs to the
predicate consist of (a) those inputs in the interior of the
separated regions, which can by identified first by computing the
residue classes of their coordinates and then by identifying which of
a finite number of polytopes (given by intersections of half-spaces
with rational coordinates) they appear within, and (b)
those inputs that lie in some slab.  The first class is
semilinear as identification of a residue class and identification
of membership in a particular polytope are both expressible in
Presburger arithmetic.  The second class is then shown to be
semilinear by induction on dimension, with the base case of dimension
0 being the trivially semilinear case of a single point.
(Section~\ref{section-proof-of-general-theorem}.)
\end{enumerate}


\section{Proof of the Semilinearity Theorem: Details}
\label{section-big-proof-details}

The proof of Theorem~\ref{theorem:general} is delayed to
Section~\ref{section-proof-of-general-theorem}.
First we describe the process of separating inputs in more detail.


\subsection{Separating Intermixed Inputs}
\label{section:intermixed}

Let $\predicate : \Population(\inputsymbols) \rightarrow \{0, 1\}$ 
be a predicate that admits finite coset coverings. 
Let $G$ be a subgroup of 
$\Z^\inputsymbols$ and let $H = h + G$ be a coset of $G$.
Take $\{(a_i, M_i)\}_{i\in I}$ and $\{(b_j, N_j)\}_{j\in J}$ to be 
finite monoid-coset covers of $\predicate^{-1}(0) \cap H$ 
and $\predicate^{-1}(1) \cap H$ respectively. 
Assume without loss of generality that $M_i, N_j \subseteq G$ by 
intersecting with $G$ if necessary. 
Define $K(i,j) := \Z (M_i \cap N_j)$, 
the group generated by the intersection of $M_i$ and $N_j$.


\begin{lemma}
\label{lemma:grouppairsep}
For all $i \in I$ and $j \in J$, no coset of $K(i,j)$ intersects 
both $a_i + M_i$ and $b_j + N_j$.
\end{lemma}
\begin{proof}
If we suppose the lemma is false, then there exist 
$x \in a_i + M_i$ and $x' \in b_j + N_j$ 
such that $(x - x') \in K(i,j)$. 
Because $M_i \cap N_j$ is a monoid, we can rewrite $x - x'$ as a difference 
$y' - y$ where $y \in M_i$ and $y' \in N_j$. 
Thus $x + y = x' + y'$. 
The former is in $a_i + M_i$, whereas the latter is in 
$b_j + N_j$. This contradicts the fact that $a_i + M_i$ 
and $b_j + N_j$ are disjoint (they have different predicate values).
\end{proof}

Define 
\[ K := \bigcap_{K(i,j) \textup{ has finitely many cosets in } G} K(i,j). \]
In addition to having finitely many cosets in $G$, 
the group $K$ inherits 
the relevant properties of the groups $K(i,j)$ included in its defining 
intersection.
\begin{lemma}
\label{lemma:groupallsep}
For all $i \in I$ and $j \in J$ such that $K(i,j)$ has finitely many cosets 
in $G$, no coset of $K$ intersects both $a_i + M_i$ and $b_j + N_j$.
\end{lemma}
\begin{proof}
Since $K(i,j) \supseteq K$, each coset of $K$ is contained in a coset
of $K(i,j)$.  The result follows from Lemma 22.
\end{proof}

Since all group cosets are semilinear, we can use membership 
in the cosets of $K$ to separate those pairs of monoid cosets that 
are intermixed. For the others, we have to take another approach.


\subsection{Separating with Hyperplanes}
\label{section:hyperplanes}

Let $g_1 + K, \ldots, g_n + K \subseteq G$ be the cosets of $K$ in $G$. 
For each $\ell \in \{1,\ldots,n\}$ define 
$H_\ell := h + g_\ell + K$. Note that $\cup_{\ell} H_\ell = H$.
If two monoid cosets $a_i + M_i$ and $b_j + N_j$ meet
both meet some coset $h + (g_\ell + K) \subseteq H$,
we show it is possible to separate
$H_\ell \cap (a_i + M_i)$ from $H_\ell \cap (b_j + N_j)$ with
a hyperplane (except for a set of lower dimension).
Define 
$I_\ell := \{i \in I \mid (a_i + M_i) \cap H_\ell \neq \emptyset\}$ and 
$J_\ell := \{j \in J \mid (b_j + N_j) \cap H_\ell \neq \emptyset\}$. 
If $i \in I_\ell$ and $j \in J_\ell$, then $H_\ell$, which is a coset
of $K$, intersects both $a_i+M_i$ and $b_j+N_j$, so $K(i,j)$
has infinitely many cosets in $G$, by 
Lemma~\ref{lemma:groupallsep}.

At this point we turn to the methods of geometry. 
We pass from groups to vector spaces by working in the vector space closure 
$\R G$ of $G$ in $\R^\inputsymbols$. Instead of monoids, we consider 
the convex cones they generate: the set of nonnegative linear combinations of 
monoid elements. 
We connect the geometry to the algebra by observing that 
$K(i,j)$ has finitely many cosets in $G$ 
if and only if $\R K(i,j)$, the vector space closure of $K(i,j)$, 
is all of $\R G$. 
Thus if two monoid cosets are not intermixed, their intersection of their 
associated monoids has 
strictly smaller dimension than $G$.
This allows us to separate these monoids with a hyperplane.

Formally, given sets of vectors $U$ and $U'$, a set of nonzero vectors 
$V$ \concept{distinguishes $U$ and $U'$} 
if for all $u \in U$ and $u' \in U'$, there exists $v \in V$
such that $(u \cdot v)(u' \cdot v) \le 0$; 
that is, either one dot product is zero or 
one is negative and the other is positive.
Define $\widehat M_\ell := \bigcup_{i\in I_\ell} M_i$ and 
$\widehat N_\ell := \bigcup_{j\in J_\ell} N_j$.
The goal of this subsection is to show the existence of a finite set of 
vectors $V$ that distinguishes $\widehat M_\ell$ from $\widehat N_\ell$. 
The main tool we use is the Separating Hyperplane Theorem 
from convex geometry.
Note that $\interior U$ denotes the interior of $U$.

\begin{theorem}
\label{theorem:SHT}
(Separating Hyperplane Theorem~\cite{lay:convex_sets}) 
If $U$ and $U'$ are convex subsets of $\R^d$ with nonempty
interiors
such that $\interior U \cap \interior U' = \emptyset$, 
then there exists $v \in \R^d$ such that $u \cdot v \le 0$ for all 
$u \in U$ and $u' \cdot v \ge 0$ for all $u' \in U'$.
\end{theorem}

Unfortunately, $M_i$ and $N_j$ are not convex. 
Thus we are forced to consider the convex cones that they generate. 
We need Carath\'eodory's theorem, another result from 
convex geometry, to verify the seemingly trivial fact that 
the intersection of these cones lies in a proper vector subspace of $\R G$. 
\begin{theorem}
\label{theorem:cara}
(Carath\'eodory's Theorem~\cite{lay:convex_sets}) 
For any set of vectors $Y \subseteq \R^d$, if $x \in \R_+ Y$, then 
there exists a linearly independent subset $Y' \subseteq Y$ such that 
$x \in \R_+ Y'$.
\end{theorem}

\begin{lemma}
\label{lemma:notfulldim}
For all $i \in I_\ell$ and $j \in J_\ell$, the set 
$Z = \R_+ M_i \cap \R_+ N_j$ 
is contained in a proper vector subspace of $\R G$. 
\end{lemma}
\begin{proof}
Suppose to the contrary. Then the vector subspace $\R Z$ is all of $\R G$ 
and has a basis $\{z_1,\ldots,z_d\} \subseteq Z$. 
By perturbing this basis to have rational coordinates, we 
can assume without loss of generality that $z_s \in \Q G$ as well for all 
$s \in \{1,\ldots,d\}$. 

By Theorem~\ref{theorem:cara}, for each $s$, there is 
a linearly independent set $Y_s \subseteq M_i$ such that $z_s \in \R_+ Y_s$. 
When we write each $z_s$ as its unique 
linear combination of elements in $Y_s$, we see by linear algebra over 
$\Q G$ that in fact $z_s \in \Q_+ M_i$. 
Repeating this argument with $N_j$ yields that 
$z_s \in \Q_+ M_i \cap \Q_+ N_j$. 
We clear denominators to find numbers $m_s$ such that 
$m_s z_s \in M_i \cap N_j$. 
The set $\{m_s z_s\}$, however, is a basis for $\R G$, which 
contradicts the fact that $\R K(i,j)$ is not all of $\R G$.
\end{proof}

\begin{lemma}
\label{lemma:hyperplane}
For all $i \in I_\ell$ and $j \in J_\ell$,
there exists a nonzero vector $v$ 
such that $\{v\}$ distinguishes $I_\ell$ and $J_\ell$. 
\end{lemma}
\begin{proof}
If either $M_i$ or $N_j$ is contained in a proper vector subspace
of $\R G$, then take $v$ to be normal to that subspace.

Otherwise, consider the sets $U := \R_+ M_i$ and 
$U' := \R_+ N_j$. 
The intersection of their interiors is both open and 
contained in a proper vector subspace of $\R G$. 
Thus $U$ and $U'$ have no interior point in common. 
Therefore, by Theorem~\ref{theorem:SHT} there exists 
$v \in \R G$ such that for all $u \in U$, we have
$u \cdot v \le 0$; and for all $u' \in U'$, 
we have $u' \cdot v \ge 0$. It follows that $\{v\}$ distinguishes 
$M_i$ and $N_j$. 
\end{proof}

To obtain a set of vectors that distinguish $\widehat M_\ell$ and 
$\widehat N_\ell$, we put together all of the individual distinguishing 
vectors. 

\begin{lemma}
\label{lemma:classifiersexist}
For all $1 \le \ell \le n$, there exists a finite set of vectors 
$V$ that distinguishes $\widehat M_\ell$ and 
$\widehat N_\ell$.
\end{lemma}
\begin{proof}
The set $V := \{v(i,j) \mid i \in I_\ell \textup{ and } j \in J_\ell\}$ 
distinguishes $\widehat M_\ell$ and $\widehat N_\ell$, where $v(i,j)$ 
is the vector from Lemma~\ref{lemma:hyperplane} such that $\{v(i,j)\}$ 
distinguishes $M_i$ and $N_j$. 
\end{proof}

Combining results in this section and the previous one, 
we have some powerful criteria 
for determining the predicate value of an input. 

\begin{lemma}
\label{lemma:transfer}
Let $V$ be a set of vectors that distinguishes $\widehat M_\ell$ and 
$\widehat N_\ell$. 
Suppose $x \in \widehat M_\ell$ 
and $y \in H_\ell$ are such that 
for all $j \in J_\ell$ and $v \in V$, 
we have $(x \cdot v)((y - b_i) \cdot v) > 0$. 
Then $\predicate(y) = 0$.
\end{lemma}
\begin{proof}
Suppose to the contrary that $\predicate(y) = 1$. Then there exists $j \in J_\ell$ 
such that $(y - b_j) \in N_j$. The set $V$, however, fails to distinguish 
$x$ and $y - b_j$, which is a contradiction.
\end{proof}

Clearly, this lemma has an analogous counterpart that establishes 
sufficient conditions for $\predicate(y) = 1$.


\subsection{Achieving Rationality}
\label{section:rationality}

The chief obstacle yet to be overcome is that a distinguishing set of 
vectors might include vectors with irrational coordinates. In order to 
rule out predicates like $\predicate(r,s) := [r < (\sqrt{2})s]$, we need to 
show that we can always distinguish $\widehat M_\ell$ and $\widehat N_\ell$ 
by vectors with integral coordinates. 
We must use for a second time the fact that $\predicate$ admits finite coset covers. 

\begin{lemma}
\label{lemma:rationality}
For all $\ell$ there exists a finite set of vectors 
from $\Z^\inputsymbols$ 
that distinguishes $\widehat M_\ell$ and $\widehat N_\ell$.
\end{lemma}

\begin{proof}
Since we can clear denominators, it is enough to find 
a set of vectors in $\Q^\inputsymbols$ that distinguishes $\widehat
M_\ell$ and $\widehat N_\ell$. 
By Lemma~\ref{lemma:classifiersexist} there exists a finite set of vectors 
$V$ that distinguishes $\widehat M_\ell$ and $\widehat N_\ell$. 
Assume that $V \setminus \Q^\inputsymbols$ has minimum cardinality,
that is, as few vectors in $V$ have irrational components as possible.
To 
avoid a special case later, we assume without loss of generality 
that $V$ contains the standard basis $\inputsymbols$ of $\R^\inputsymbols$. 
We show that this set $V$ contains only vectors in $\Q^\inputsymbols$.

Suppose to the contrary that there exists 
$v \in V \setminus \Q^\inputsymbols$. 
By assumption, the set $V' := V \setminus \{v\}$ cannot distinguish 
$\widehat M_\ell$ and $\widehat N_\ell$. 
Thus there exist $i \in I_\ell$ and $j \in J_\ell$ and 
$x \in M_i$ and $y \in N_j$ such that 
for all $v' \in V'$ we have $(x \cdot v')(y \cdot v') > 0$.

We consider two cases. In the first case, for all choices of $x$ and $y$ 
we have $(x \cdot v)(y \cdot v) = 0$. Then at least one vector in each 
problem pair is normal to $v$. By linear algebra over $\Q^\inputsymbols$, 
there exists a vector $v' \in \Q^\inputsymbols$ such that 
if $w \in G$ is normal to $v$, 
then $w$ is normal to $v'$. Thus $V' \cup \{v'\}$ distinguishes 
$\widehat M_\ell$ and $\widehat N_\ell$, which is a contradiction, 
since $(V' \cup \{v'\}) \setminus \Q^\inputsymbols$ has fewer elements 
than $V \setminus \Q^\inputsymbols$. 

In the second case, we have $x$ and $y$ such that 
$(x \cdot v')(y \cdot v') > 0$ for all $v' \in V'$ and 
$(x \cdot v)(y \cdot v) < 0$. 
Assume without loss of generality that 
$x \cdot v < 0 < y \cdot v$ by taking $-v$ instead of $v$ if necessary. 
Let $\Omega := \{w \in \R G \mid v' \cdot w > 0 \textup{ for all } 
v' \in V \textup{ and } w(q) > 0 \textup{ for all } q \in \inputsymbols \}$.
Clearly $\Omega$ is an open set. Also, $y \in \Omega$, since by the 
assumption that $\inputsymbols \subseteq V$ we have $y(q) > 0$ 
for all $q \in \inputsymbols$. 
Given that $\Omega$ is a nonempty open set, 
we can extend $x,y$ to a basis $w_1 = x, w_2 = y, \ldots,w_m$ such that 
$w_s \in \Omega$ for all $s \ge 2$. 
By perturbing each $w_s$ slightly to have rational coordinates 
and clearing denominators, 
we assume without loss of generality that $w_s \in \Omega \cap K$. 
There exists $s$ such that $(w_s \cdot v)/(w_1 \cdot v)$ is irrational, 
since otherwise some scalar multiple of $v$ belongs to $\Q^\inputsymbols$. 

In consequence $\N(x \cdot v) + \N(w_s \cdot v)$ is dense in $\R$, 
so we can find sequences of positive integers $m_t$ and $m_t'$ such that
\[m_t(x \cdot v) + m_t'(w_s \cdot v)\]
is a negative monotone increasing sequence of real numbers approaching
$0$ as $t$ approaches infinity.
For all $v' \in V$, the points $(m_t x + m_t' w_s)_{t\ge 1}$ lie 
on the same side of the hyperplane normal to $v$, 
and as a sequence they approach the hyperplane normal to $v$ 
arbitrarily closely. 
By another application of Higman's Lemma, there is an increasing function 
$f$ such that $(m_{f(t)}, m_{f(t)}')$ is an increasing sequence.

Let $z \in (a_i + M_i) \cap H_\ell$.
There exists some constant $c \ge 0$ such that for each $r$, 
the points $((z + (c+m_{f(t)}) x + m_{f(t)}' w_s) - b_r)_{t\ge 1}$ 
all lie on the same side of each hyperplane as $x$. 
It is easily verified that each of these points belongs to $H_\ell$. 
Thus by Lemma~\ref{lemma:transfer}, 
$\predicate(z + (c+m_{f(t)}) x + m_{f(t)}' w_s)$ is constantly $0$. 
Apply the pumping lemma (Lemma~\ref{lemma:pumping}) again to
these inputs to
obtain a finite cover. Then there exist $t_1 < t_2$ such that 
$(m_{t_2} - m_{t_1}) x + (m_{t_2}' - m_{t_1}') w_s \in P$, where the monoid 
coset $(z + m_{t_1} x + m_{t_2} w_s) + P$ belongs to the cover.
Pumping $z + (c+m_{t_1}) x + m_{t_1}' w_s$ by a sufficiently large multiple 
of $(m_{t_2} - m_{t_1}) x + (m_{t_2}' - m_{t_1}') w_s$ 
yields an element $z'$ such that $\predicate(z') = 0$, but for each $r$, 
we have that $z' - a_r$ is on the same side of each hyperplane as $w_s$, 
which contradicts Lemma~\ref{lemma:transfer}. 
\end{proof}


\subsection{Proof of Theorem~\ref{theorem:general}}
\label{section-proof-of-general-theorem}

We can now prove Theorem~\ref{theorem:general} by
induction on the dimension of $G$ (the
cardinality of the largest linearly independent subset of $G$.)

\begin{proof}
If the dimension of $G$ is zero, then $H$ is a single point
and the result holds.
If the dimension of $G$ is greater than zero,
then by Lemmas~\ref{lemma:groupallsep} 
and \ref{lemma:rationality}, there exist a group $K$
and finite distinguishers $V_\ell \subseteq \Z^\inputsymbols$ for all 
cosets $H_\ell$ of $K$ in $H$. 

For each distinguishing 
vector $v \in V_\ell$, consider the set of points $x \in H_\ell$ such that 
it is not that the case that the following numbers are either all negative 
or all positive: 
$(x-a_i)\cdot v$ for $i \in I_\ell$ and $(x-b_j)\cdot v$ for $j \in J_\ell$. 
This set has finite width in the direction of $v$. Thus it can be written 
as the union of finitely many cosets of a group of smaller dimension. 
The key to the induction is that any point not in the union of these sets 
over the different $x$ satisfies 
the hypotheses of one of the variants of Lemma~\ref{lemma:transfer}. 

Define the sets 
\begin{align*}
B & := \bigcup_\ell \{x \in H_\ell \mid \exists 
y \in \widehat N_\ell \textup{ such that } ((x-b_j)\cdot v)(y\cdot v) > 0 
\textup{ for all } j \in J_\ell \textup{ and } v \in V_\ell \} \\
B' & := \bigcup_\ell \{x \in H_\ell \mid \not\exists 
y \in \widehat M_\ell \textup{ such that } ((x-a_i)\cdot v)(y\cdot v) > 0 
\textup{ for all } i \in I_\ell \textup{ and } v \in V_\ell \}.
\end{align*}
By Lemma~\ref{lemma:transfer} we have 
$B \subseteq \predicate^{-1}(1) \subseteq B'$. 
It is not difficult to verify that $B$ and $B'$ are semilinear. Moreover, 
$B'\setminus B$ is a union of finitely many group cosets of dimension 
less than $G$. It follows by inductive hypothesis 
that $\predicate^{-1}(1) \cap (B'\setminus B)$ is 
semilinear, and thus that $\predicate$ itself is semilinear. 
\end{proof}


\section{Characterizations of the One-Way Models}
\label{section:one-way-char}

In this section, we complete the characterizations of the one-way
models in Figure~\ref{fig:power} by showing that they are limited
to the indicated power.


\subsection{Immediate and Delayed Transmission}
\label{section:immdeltrans-upper}


We first consider the immediate and delayed transmission models to
prove the following converse 
to Theorem~\ref{theorem:immediatedelayedtrans-lower}.

\begin{theorem}
\label{theorem:immediatedelayedtrans-upper}
Let $\predicate$ be a predicate that is stably
computable by an immediate or delayed
transmission protocol.
Then $\predicate$ is in $\classSLIN \cap \classCoreMOD$.
\end{theorem}

Thus, for example, the comparison predicate, true if
there are more $a$'s than $b$'s in the input,
is not stably computable in the immediate or delayed
transmission model, because it is not in $\classCoreMOD$.
Intuitively, the weakness of delayed transmission compared
with queued transmission is in the inability of a receiver
to refuse messages temporarily; in effect, by delivering
a number of copies of some message to a receiver and
and forcing it back into the same state, we cause the
whole collection of messages to ``disappear'' from the computation.


Because a delayed transmission protocol is a special case
of a queued transmission protocol and an immediate transmission
protocol is a special case of a two-way protocol, 
Corollary~\ref{corollary:two-way-queued-char} shows that
$\predicate$ is in $\classSLIN$.
To see that $\predicate$ is also in $\classCoreMOD$ requires
some preliminary lemmas.
Consider any delayed or immediate transmission protocol
that stably computes $\predicate$, and consider 
any nonempty subset $\inputsymbols'$ of $\inputsymbols$.

To unify the treatment of the two kinds of protocols,
we assume that in a delayed transmission protocol, the
sender simply sends its current state; thus, we need
not distinguish between states and messages in our description.
Formally, in this case the set of messages consists of a disjoint
copy of the set of states, distinguished by the phrases ``the state
$q_i$'' and ``the message $q_i$.''
In the case of immediate transmission, an interaction
$(p,q) \mapsto (\delta_1(p),\delta_2(p,q))$ is 
described as the sender sending message $p$ and going to state
$\delta_1(p)$, and the receiver receiving message $p$ and going
to state $\delta_2(p,q)$.

We let $Q'$ denote the set of states that appear
in configurations reachable from inputs that contain only symbols
from $\inputsymbols'$.
For states $q, q' \in Q'$ of the  protocol, we say that
$q$ \concept{can reach} $q'$ if there is some
finite sequence of messages (from $Q'$) sent and received by
an agent in state $q$ that enable it to enter state $q'$.
Formally, $q$ can reach $q'$ if there is a sequence of states
$q=q_1, q_2, \ldots, q_k=q'$ such that, for $2\leq i \leq k$,
either $q_i = \delta_s(q_{i-1})$ or $q_i = \delta_r(p,q_{i-1})$ for
some message $p \in Q'$.


The following lemma shows that, for any message $p$, a sufficient
number of copies 
of $p$ can be ``absorbed'' by an agent in some state $q$, returning that
agent to state $q$.

\begin{lemma}
\label{lemma-absorption}
Fix any message $p\in Q'$.  Define $f_p(q) = \delta_r(p,q)$.
There exists a state $q\in Q'$ and a positive integer $n$ such that
$f_p^{(n)}(q)=q$.
\end{lemma}

\begin{proof}
For any state $r \in Q'$, 
the sequence of states $r, f_p(r),  f_p^{(2)}(r), \ldots$
in $Q'$ must be eventually periodic of some period $n$.
Choosing $q$ to be any state in the periodic part,
we see that $f_p^{(n)}(q) = q$.
\end{proof}

If $n$ is a positive integer and $q_1, q_2 \in Q'$ are states,
we say that $q_1$ and $q_2$ are $n$-\concept{substitutable}
if there exist an input $x$ containing only symbols from $\inputsymbols'$
and a configuration $b$ of states,
such that the configurations $c = nq_1 + b$ and $d = nq_2 + b$ 
are reachable from $I(x)$.
We note that if $q_1$ and $q_2$ are $n$-substitutable,
then they are $mn$-substitutable for any positive integer $m$.
We define two states $q_1, q_2 \in Q'$ to be \concept{substitutable}
if there is some positive integer $n$ such that they
are $n$-substitutable.
Substitutability is reflexive and symmetric by definition;
we now show it is transitive.

\begin{lemma}
\label{lemma-substitutability-transitive}
Let $q_1, q_2, q_3 \in Q'$ be states and assume that
$q_1$ and $q_2$ are substitutable and $q_2$ and $q_3$
are substitutable.
Then $q_1$ and $q_3$ are substitutable.
\end{lemma}

\begin{proof}
For some input $x$ containing only symbols from $\inputsymbols'$, 
configuration $b$ of states and integer
$n$, the configurations $c = nq_1 + b$
and $d = nq_2 + b$ are reachable from $I(x)$.
Similarly, for some input $y$ containing only symbols from $\inputsymbols'$, 
configuration $b'$ of states
and positive integer $m$, the configurations $c' = mq_2 + b'$
and $d' = mq_3 + b'$ are reachable
from $I(y)$.

Consider the input $z = mx + ny$, which contains only input symbols
from $\inputsymbols'$.
From $I(z)$ there is a computation reaching $mc + nc'$, which
is equal to $mnq_1 + (mnq_2 + mb + nb')$, and also a computation
reaching $md +nd'$, which is equal to $mnq_3 + (mnq_2 + mb + n b')$.
Thus $q_1$ and $q_3$ are $mn$-substitutable.
\end{proof}

Now we show that reachability implies substitutability.

\begin{lemma}
\label{lemma-reachable-implies-substitutable}
If $q_1, q_2 \in Q'$ are states such that
$q_1$ can reach $q_2$ then $q_1$ and $q_2$ are substitutable.
\end{lemma}

\begin{proof}
Because substitutability is transitive, it suffices to
consider the case in which $q_2$ is reachable in one step from $q_1$.
For every state $q \in Q'$, we define $c_q$ to be
any configuration of states
containing $q$ that is reachable from an
input containing only symbols from $\inputsymbols'$.
We consider two cases.

If $q_1$ reaches $q_2$ by a send step,
then by Lemma~\ref{lemma-absorption} we may choose a state $q_3 \in Q'$
and a positive integer $n$ such that 
$f_{q_1}^{(n)}(q_3) = q_3$.  (Intuitively, this means
an agent in state $q_3$ will
be back in $q_3$ after receiving $n$ copies of message $q_1$.)
We choose $b = n(c_{q_1} - q_1) + c_{q_3}$, and let
$c = nq_1 + b$, so that $c  = nc_{q_1} + c_{q_3}$.
Then $c$ is reachable from $I(x)$ for some input $x$ containing only
symbols from $\inputsymbols'$.

And from $c$ we may have each of $n$ agents in state $q_1$
transmit a message (and reach state $q_2$) to a single agent
in state $q_3$, which will again be in state $q_3$ after
receiving the $n$ copies of $q_1$.
Formally, it is easy to see by induction on $j$ that $c \reaches
(n-j)q_1 + jq_2 +b -q_3 +f_{q_1}^{(j)}(q_3)$ for $0\leq j \leq n$.
Taking $j=n$, we have that 
$d = nq_2 + b$ is reachable from $c$, and therefore from $I(x)$, 
so $q_1$ and $q_2$ are substitutable.

Suppose $q_1$ reaches $q_2$ by a receive step, say of message $q_3 \in Q'$.
Let $q_4$ be the state reached from state $q_3$ after a send.
Then by Lemma~\ref{lemma-absorption}, there is a state $q_5 \in Q'$ and
a positive integer $n$ such that 
$f_{q_3}^{(n)}(q_5) = q_5$.
(This means that after receiving $n$ copies of
message $q_3$ an agent that starts in state $q_5$ will be back in
state $q_5$.)
We choose 
\[b = n(c_{q_1} - q_1) + n(c_{q_3} - q_3) + nq_4 + c_{q_5},\] 
and and let
\[c = nq_1 + b = nc_{q_1} + n(c_{q_3} - q_3) + nq_4 + c_{q_5}.\]
Now $c' = nc_{q_1} + nc_{q_3} + c_{q_5}$ is reachable from $I(x)$ for 
some input $x$ containing only symbols from $\inputsymbols'$.
It is easy to see by induction on $j$ that, for $0\leq j\leq n$,
$c' \reaches nc_{q_1} + nc_{q_3} -jq_3 +j q_4 +c_{q_5} - q_5
+f_{q_3}^{(j)}(q_5)$ by a sequence of steps in which $j$ 
agents in state $q_3$ send a message (and go to state $q_4$)
to a single agent that started in state $q_5$.
Taking $j=n$, we see that 
$c' \reaches nc_{q_1} + nc_{q_3} -nq_3 +n q_4 +c_{q_5} = c$.
Also, $d = nq_2 + b = n(c_{q_1}-q_1)+nq_2+n(c_{q_3}-q_3)+nq_4+c_{q_5}$
is reachable from $c'$ by a computation 
in which 
we pair up $n$ agents in state $q_3$ with $n$ agents in state $q_1$
and in each of the $n$ pairs have the agent in state $q_3$ send a message to
the agent in state $q_1$, transforming the sender into state $q_4$ and
the receiver into state $q_2$.
Thus, in this case also, $q_1$ and $q_2$ are substitutable.
\end{proof}


Thus we have the following pumping lemma for predicates stably
computable by immediate and delayed transmission protocols.

\begin{lemma}
\label{lemma-transmission-pumping}
For every input symbol $\sigma \in \inputsymbols'$ 
there exist positive integers $k$ and $n$ such that 
for every input $y \in \Population(\inputsymbols)$ 
that is $k$-rich with respect to $\inputsymbols'$,
$y$ and $y + n\sigma$ are both accepted or
both rejected.
\end{lemma}

\begin{proof}
Let $k_0$ be the constant in Lemma~\ref{lemma:trunc-stable}
such that $k_0$-truncates are sufficient to determine membership
in the set $\stable$ of output stable configurations for the protocol.
We choose $k_1$ to be $|Q'|(k_0 - 1) + 1$.

Let $\sigma$ be any input symbol from $\inputsymbols'$ 
and let $q = \iota(\sigma)$, the initial state associated with $\sigma$.
Let $R \subseteq Q'$ 
be the set of states that $q$ can reach, and consider any $q' \in R$.
By Lemma~\ref{lemma-reachable-implies-substitutable}, $q$ and $q'$
are $n_{q'}$-substitutable for some positive integer $n_{q'}$.
Let $n$ be the least common multiple of the $n_{q'}$ for all
$q' \in R$; $q$ and $q'$ are $n$-substitutable.
That is, there exists an input $x_{q'}$ containing only symbols
from $\inputsymbols'$ and
a configuration of states $b_{q'}$ such that 
the configurations  $c_{q'} = nq + b_{q'}$ 
and $d_{q'} = nq' + b_{q'}$ are
reachable from $I(x_{q'})$.

Let $x$ be the sum of all $x_{q'}$ over $q' \in R$.
We choose $k_2$ to be the maximum multiplicity
of any input symbol in $x$.
Let $k = \max(k_1,k_2)$ and let $y$ be any 
input that is $k$-rich with respect to $\inputsymbols'$.
Then $y \ge x$ and $y$ contains only symbols from $\inputsymbols'$.

Consider a computation from $I(y)$ that first takes $I(x)$ to the
sum  $s = \sum_{q' \in R} c_{q'}$
and then proceeds to an output stable configuration $e$.
Since $y$ is $k$-rich with respect to $\inputsymbols'$, 
there are at least $k\ge k_1$ agents.
By the pigeonhole principle, some state $q'$ appears with multiplicity
at least $k_0$ in $e$.
Note $q' \in R$.

Now from $I(y + n\sigma) = I(y) + nq$ we consider a computation that
takes $I(x - x_{q'})$ to $s - c_{q'}$ and $I(x_{q'})$ to $d_{q'}$,
giving
\[I(y - x) + nq + (s - (nq + b_{q'})) + nq' + b_{q'} = I(y-x) + s + nq'.\]
Now we continue by running $I(y-x) + s$ to $e$, giving $e + nq'$,
which is output stable because $\tau_{k_0}(e) = \tau_{k_0}(e + nq')$.
Thus $y$ and $y+n\sigma$ are both accepted or both rejected.
\end{proof}

Now we can conclude the proof of 
Theorem~\ref{theorem:immediatedelayedtrans-upper}.
\begin{proof}
For each alphabet symbol $\sigma \in \inputsymbols'$, 
by Lemma~\ref{lemma-transmission-pumping}
there exist integers $k_{\sigma}$ and $n_{\sigma}$ 
such that for all inputs $y$ that are $k_{\sigma}$-rich
with respect to $\inputsymbols'$,
$y + n_{\sigma}\sigma$ is accepted if and only if $y$ is accepted.
Let $k$ be the maximum of the $k_{\sigma}$'s over $\sigma \in \inputsymbols'$.
Then the acceptance or rejection of an input $y$ that is
$k$-rich with respect to $\inputsymbols'$ depends
only on the values of the number of occurrences of $\sigma$
modulo $n_{\sigma}$, which implies that
$\predicate$ is $k$-similar with respect to $\inputsymbols'$ to
a predicate in in $\classMOD$.
Because $\inputsymbols'$ was an arbitrary nonempty subset of $\inputsymbols$,
this shows that $\predicate \in \classCoreMOD$.
\end{proof}


\subsection{Immediate Observation}
\label{immediate-observation}

In the immediate observation model, transitions are
of the form $(p,q) \mapsto (p,q')$ and
there is no multiset of undelivered messages.
We have shown in Theorem \ref{theorem:immediateobs-lower}
that for any constant $k$, an immediate observation
protocol can count the number of copies of each input symbol up to $k$,
However, this is also the extent of its power.


\begin{theorem}
\label{theorem:immediate-obs-upper}
Let $\predicate$ be a predicate that is stably
computable by an immediate observation protocol.
Then $\predicate \in \classCOUNT_*$.
\end{theorem}
\begin{proof}
Let $k$ be the constant in Lemma~\ref{lemma:trunc-stable}
such that $k$-truncates are sufficient to determine membership
in the set $\stable$ of output stable configurations.
Assume that $x$ is an input such that some
input symbol $\sigma$ occurs
with multiplicity at least $k'=|Q|(k-1) + 1$ in $x$.
Let $c_0,c_1,c_2, \ldots,c_m$ be an execution from the initial
configuration $c_0=I(x)$ to an
output stable configuration $c_m$.  Then, $c_i=c_{i-1}-p_i+p_i'$ for
some $p_i,p_i'\in Q$.
We define $s_i(j)$ inductively to represent the state of the $j$th
agent with input $\sigma$ in $c_i$, as follows.
Let
$s_0(j)=\iota(\sigma)$.
If the multiset $\{s_{i-1}(j) : 1\leq j\leq k'\}$ is contained in
$c_i$, then $s_i=s_{i-1}$.  Otherwise, there is some $\jhat$ such
that $s_{i-1}(\jhat)=p_i$, and we define 
$s_i(j) = 
\begin{cases} 
p_i' &\mbox{if }j=\jhat\\
s_{i-1}(j)&\mbox{otherwise.}
\end{cases}$

By the pigeonhole principle, the multiset 
$\{s_m(j) \mid 1\leq j\leq k'\}$ 
contains some state $q'$ with multiplicity at least $k$.  
Choose $j'$ such that $s_m(j')=q'$.  
We introduce a
``clone'' of this agent into the execution.  
Formally, we show by induction that 
$I(x+\sigma) \reaches c_i + s_i(j')$.  
When $i=0$, $I(x+\sigma) = c_0+s_0(j')$.
Assume $I(x+\sigma) \reaches c_{i-1} + s_{i-1}(j')$.  
If $q_i(j') = s_{i-1}(j')$, then the claim is clearly true.  
Otherwise,
$c_{i} + s_{i}(j') = c_{i-1} -p_i+ 2p_i'$, 
which is reachable from
$c_{i-1}+s_{i-1}(j') = c_{i-1}+p_i$ by having two agents in state $p_i$
change to state $p_{i}'$ by observing some state in $c_{i-1}-p_i$.

Thus, $I(x + \sigma) \reaches c_m + q'$.
Because the multiplicity of $q'$ in $c_m$ is at least $k$,
$\tau_k(c_m) = \tau_k(c_m+q')$.
Therefore, since $c_m$ is output stable, so is $c_m + q'$.
Moreover, $c_m + q'$ has the same output as $c_m$, so
$\predicate(x) = \predicate(x + \sigma)$.
Because this holds of any input symbol $\sigma$ of
multiplicity at least $k'$ in $x$, we
have $\predicate \in \classCOUNT_{k'}\subset \classCOUNT_*$.
\end{proof}


We may also consider a variant of the immediate observation
model in which
an agent may interact with itself, that is,
a rule $(p, p) \mapsto (p,q)$ can be applied
to a single agent in state $p$ to change it
to state $q$.
This is the model \concept{with mirrors}.
The two versions of the model are equal in power,
as they can simulate
each other, using bit-flipping protocols
to avoid or permit self-interactions.

\begin{theorem}
\label{theorem:immediateobservationmirrors}
The same predicates are stably computable by immediate observation
protocols in the models with and without mirrors.
\end{theorem}

\begin{proof}
We may assume that $n$, the population size, is at least
$3$, since any predicate on $3$ or fewer agents is computable
in either model; this case can be detected and handled
separately in either kind of protocol.

Given an immediate observation protocol that stably
computes a predicate $\predicate$ in the model without mirrors,
we can derive another protocol that stably
computes $\predicate$ in the model with mirrors.
For every old state $q$ we introduce a new state $q'$
and for every old transition $(p,q) \mapsto (p,r)$, where
$p \ne q$, we introduce variants of the transition with
all combinations of primed and unprimed states $p$, $q$, and $r$.
We replace every old transition $(p,p) \mapsto (p,q)$
with four new transitions: $(p,p) \mapsto (p,p')$ and
$(p',p') \mapsto (p',p)$, as well as $(p,p') \mapsto (p,q)$
and $(p',p) \mapsto (p',q)$.
In the mirrored model, this allows an agent to flip
back and forth between $p$ and $p'$ by itself,
but to leave these two states, it must interact
with someone other than itself (witnessed by having
different ``primation'').

For the other direction, given an immediate
observation protocol in the model with mirrors
that stably computes a predicate $\predicate$, 
we can again introduce primed and unprimed versions
of each state.
For each ordered pair of states $p$ and $q$, we have
rules $(p,q) \mapsto (p,q')$ and $(p,q') \mapsto (p,q)$,
which allow a state to flip between primed and unprimed
as long as there is at least one unprimed state.
Because the input states are unprimed, these rules cannot
eliminate the last unprimed state.
In addition, for every rule $(p,q) \mapsto (p,r)$ with $p \neq q$
in the protocol with mirrors, there is a rule $(p',q) \mapsto (p',r)$
in the protocol without mirrors; this allows steps involving two
agents in different states to be simulated by priming the initiator,
unpriming the responder, and taking the step.
For every rule $(p,p) \mapsto (p,r)$ in the protocol with mirrors,
there are rules $(q',p') \mapsto (q',r')$ for every original state $q$.
If a step involves two agents in state $p$, it can be simulated
by priming both agents and taking the step.
If a step involves one agent in state $p$, it can be simulated
by priming that agent (to get $p'$) and any other agent (to get $q'$)
and taking the step.
Note that this simulation can be carried out if there are at least
three agents in the population because there is at least one unprimed
agent in every configuration reachable from an input configuration.
\end{proof}



\subsection{Delayed Observation}

Here, we prove the converse of Theorem~\ref{theorem:delayedobs-lower},
showing that the delayed observation model is the weakest of the
one-way models we defined.

\begin{theorem}
\label{theorem:delayedobs-upper}
Suppose $\predicate$ is stably computed by a delayed
observation protocol.
Then $\predicate$ is in $\classCOUNT_1$.
\end{theorem}

\begin{proof}
We show that for any input $x$, if $\sigma \in x$ then
$\predicate(x+\sigma) = \predicate(x)$, which implies that $\predicate$ 
is completely determined by
the presence or absence of each input symbol and hence is in $\classCOUNT_1$.

Consider the finite graph whose nodes are configurations reachable from
$I(x)$ that contain
no messages in transit, with a directed edge from $c$ to $c'$ if 
$c \reaches c'$.
A \concept{final} strongly connected component of this graph is one from
which no other strongly connected component of the graph is reachable.
From $I(x)$ we can reach a configuration in a final strongly
connected component $\cF$ of this graph.
Let $\hat \cF$ denote all the configurations $d$, including those with
undelivered messages,  such that 
$c \reaches d$ for some $c \in \cF$.
For any configurations $d$ and $d'$ in $\hat \cF$, $d \reaches d'$
by first delivering all messages in $d$.
This implies that all configurations in $\hat \cF$ are output stable.

The set $T$ of states that occur in configurations in $\hat \cF$ 
is closed, that is, if $p, q \in T$ and
$(p,q) \mapsto (p,q')$, then $q' \in T$.
To see this, assume not.  Then, take a configuration $c$ in $\hat \cF$ that
contains $p$ 
and let an agent in state $p$ send a message, putting $p$ into
messages in transit.
Now mimic a computation from $d$ to a configuration $d'$ in $\hat \cF$
containing $q$, leaving the message $p$ undelivered.
Then deliver $p$ to an 
agent in state $q$, arriving at a configuration in $\hat \cF$
containing $q'$, a contradiction.

Now consider any $\sigma$ in $x$.
Consider an execution that begins in $I(x)$ and ends in an output
stable configuration $c$ in $\cF$.
There must be a sequence of states $q_0, q_1, \ldots, q_m$ where $q_0 =
\iota(\sigma)$, $q_m$ appears in $c$, and $(p_{i},q_{i-1})\mapsto
(p_{i},q_i)$ for some state $p_{i}$ that appeared in some configuration
during the execution.
Starting from the configuration $I(x+\sigma)=I(x)+q_0$, we can reach a
configuration $c+q_0$ that also has one additional copy of each message
$p_1, \ldots, p_{m}$ left in transit.  By delivering all of the $m$
messages to the agent in state $q_0$ in order, we reach the
configuration $c+q_m$.
Because $T$ is closed and the states of $c+q_m$ are all in $T$,
$c+q_m$ is output stable and therefore
$\predicate(x+\sigma) = \predicate(x)$.
\end{proof}

If we assume that no agent ever receives its own message,
then the power of the model increases only slightly: it
becomes possible for an agent in a unique state to observe that it
never encounters a twin, and the stably computable predicates
are $\classCOUNT_{2}$ in this case.
The proof is a straightforward extension of the
proof of Theorem~\ref{theorem:delayedobs-upper}.


\section{Local Fairness Is Weak Even with Unbounded States}
\label{section-local-fairness}

In this section,
we consider a strongly anonymous message-passing model with
the following local fairness condition: if some agent sends a
particular message $m$ infinitely often, 
then each agent
receives message $m$ infinitely often.
This model turns out to be surprisingly weak.
Even if the states of processes and the 
lengths of messages may grow without bound,
protocols in this model cannot distinguish two
multisets of inputs if the same set of values appears in each.
Since this model subsumes the one-way finite-state population protocol
models, this result
supports the choice of the stronger global fairness
condition assumed in the rest of the paper.

\begin{theorem}
\label{theorem-weak-fairness}
Let $\inputsymbols$ denote the (finite or countably infinite) 
set of possible input values.
A predicate $\predicate$ on finite nonempty 
multisets of elements from $\inputsymbols$
is stably computable in the asynchronous
message-passing model with the weak fairness condition if and only if
$\predicate$ is completely determined by the set of input values
present in the initial configuration.
\end{theorem}

\begin{proof}
We describe a protocol in which each agent eventually determines the
set of all the inputs that occur in the initial configuration,
and therefore the correct value of $\psi$.
The state of each agent is the set consisting of its initial
input value together with every value that has occurred in
a message it has received.
Whenever an agent runs, it sends its current state.
Whenever an agent receives a message, it updates its state
to be the union of its previous state and the set of values
in the message.

Clearly every message sent is a subset of the set of input values
in the initial configuration,
so there are only finitely many possible messages in each computation.
Every message sent by an agent with input value $\sigma$ contains the
element $\sigma$, and it sends infinitely many messages, so eventually
every agent receives a message containing $\sigma$.
Thus, the state of every agent eventually consists of the
set of values in the input configuration, 
and each outputs the correct value of $\predicate$.

For the converse, 
assume that we have a protocol that stably computes 
a predicate $\predicate$, and let $x$ and $x'$ be any two 
nonempty multisets of values from
$\inputsymbols$ such that the same set of values appears in each.
Let $n = |x|$ and $n' = |x'|$.  Let $c_0$ and $c_0'$ be initial
configurations corresponding to inputs $x$ and $x'$, respectively.
We construct two executions $\alpha$ and $\alpha'$ starting from $c_0$
and $c_0'$.
Let $m_1,m_2,\ldots$ be an arbitrary sequence 
of messages where every possible
message appears infinitely often.
We construct the executions $\alpha$ and $\alpha'$
in phases, where phase $i$ will
ensure that message $m_i$ gets delivered to everyone
if that message has been sent enough times.
Let $c_i$ and $c_i'$ be the configurations of $\alpha$ and $\alpha'$ 
at the end of phase $i$.  

Our goal is to prove the following claim:
for all $i\geq 0$ and for all $\sigma \in \inputsymbols$, 
the state in $c_i$ of each agent with original input $\sigma$ is
the same as the state in $c_i'$ of each agent with original
input $\sigma$.
Assume that we have constructed the first $i-1$ phases of the two
executions so that the claim is satisfied.  
Suppose we run all processes in lock step from $c_{i-1}$ and
$c_{i-1}'$ without delivering any messages.
There are two cases.

Case (i):
Eventually, after $r_i$ rounds, the run from $c_{i-1}$ will have at
least $n$ copies of $m_i$ in transit and, after $r_i'$ rounds, the
run from $c_{i-1}'$ will have at least $n'$ copies of $m_i$ in transit.
Then, the $i$th phase of $\alpha$ and $\alpha'$ is constructed by
running each agent for $\max(r_i,r_i')$ rounds without
delivering any messages, and then delivering one
copy of $m_i$ to every agent.
This ensures the claim will be true for $c_i$ and $c_i'$.

Case (ii):
Otherwise, we allow every agent to take one step without delivering
any messages.  (This clearly
satisfies the claim for $c_i$ and $c_i'$.)

It remains to show that both $\alpha$ and $\alpha'$ 
satisfy the weak fairness condition, and then
it will follow from the claim that $\predicate(x) = \predicate(x')$.
First, notice that every agent takes infinitely many steps in
$\alpha$ and $\alpha'$.
If some agent $v$ sends a message $m$ infinitely many times in $\alpha$ or
$\alpha'$, it will also be sent infinitely many times by an agent
with the same input value in the other execution (since an agent with a
particular input experiences the same sequence of events in both
executions).
Suppose $m$ is never delivered after phase $i$ to some agent $w$
in one of the two executions.  Eventually, there
will be $n$ copies of $m$ in transit in $C_j$ for some $j>i$ and $n'$
copies of $m$ in transit in $C_{j'}'$ for some $j'>i$.  Consider the
first occurrence of $m$ in the sequence $m_1, m_2, \ldots$ that comes
after $m_j$ and $m_j'$.  During the corresponding phase, $m$ will be
delivered to every agent, including $w$, a contradiction.
Thus, $\alpha$ and $\alpha'$ satisfy the weak fairness condition.
\end{proof}


\section{Conclusions and Discussion}

We have shown that the predicates stably computable by population
protocols in the all-pairs communication graph
are semilinear, answering the main open question 
in~\cite{AADFP04,AADFP06}.
This result also shows that the model of population protocols
with stabilizing inputs, introduced in \cite{AACFJP05}, is equal
in computational power to the standard model of population protocols
in the all-pairs communication graph,
resolving another open question.

Our semilinearity proof is given for an abstract model that
includes not only population protocols, but generalizations
permitting rules that replace one finite multiset of elements of
a configuration with another finite multiset of elements.
Thus, even a generalization of population protocols in which
rules may add or delete agents or involve interactions between more
than two agents will stably compute only semilinear
predicates, answering a question in~\cite{AADFP03}.

We have introduced several models of one-way communication in
population protocols: queued transmission, immediate and delayed
transmission and immediate and delayed observation, and exactly
characterized the classes of predicates stably computable in these
models in the all-pairs communication network
by natural subclasses of the semilinear predicates.
These one-way models are more closely related to asynchronous
message-passing models, and illuminate the effects of capabilities
such as the ability to refuse to accept incoming messages temporarily.

Another natural variant of the population protocol model
considers ``reversing moves'' and requires that a protocol
stably compute a predicate despite the possible presence
in an execution of a finite number of reversing moves, that is,
steps in which a rule $(p,q) \mapsto (p',q')$ is used
in reverse: $(p', q') \mapsto (p, q)$.
It is  not difficult to show that the basic protocols for
modulo and threshold predicates are resistant to such reversing
moves, and therefore that all the semilinear predicates can
be computed by protocols resistant to reversing moves.
Thus, requiring such resistance does not reduce the computational
power of population protocols in the all-pairs communication graph.

Despite promising results for population protocols in restricted
communication graphs~\cite{AACFJP05}, much more remains to be
understood about their computational power.
In particular, the computational power of one-way protocols in
restricted communication graphs has not been studied.

\section{Acknowledgments}

This research was carried out while the third author was an
undergraduate at the University of Rochester.
The authors would like to thank the reviewers of the extended
abstracts on which this paper is based~\cite{AAE06,AAER05} 
for their thoughtful comments and suggestions.


\bibliographystyle{alpha}
\bibliography{journal}


\end{document}